\documentclass[fleqn,usenatbib]{mnras}

\usepackage[T1]{fontenc}

\DeclareRobustCommand{\VAN}[3]{#2}
\let\VANthebibliography\thebibliography
\def\thebibliography{\DeclareRobustCommand{\VAN}[3]{##3}\VANthebibliography}

\usepackage{graphicx}	
\usepackage{amsmath}	
\usepackage{amssymb}	
\usepackage{mhchem}
\usepackage{listings}
\usepackage{xcolor}
\usepackage{fontawesome5}

\usepackage{newtxtext,newtxmath}

\definecolor{backcolour}{rgb}{0.94,0.97,1.0}
\definecolor{codeblue}{rgb}{0.36, 0.54, 0.66}
\lstdefinestyle{mystyle}{
    backgroundcolor=\color{backcolour},   
    commentstyle=\color{codeblue},
    breakatwhitespace=false,         
    breaklines=true,                 
    captionpos=b,                    
    keepspaces=true,                 
    numbers=left,                    
    numbersep=5pt,                  
    showspaces=false,                
    showstringspaces=false,
    showtabs=false,                  
    tabsize=2
}
\definecolor{mygrey}{RGB}{30, 30, 29}
\definecolor{mywhite}{RGB}{245, 252, 252}

\title[XRayTheSpot: X-raying Molecular Gas]{Tree-based solvers for adaptive mesh refinement code FLASH - IV: An X-ray radiation scheme to couple discrete and diffuse X-ray emission sources to the thermochemistry of the interstellar medium}

\author[Gaches et al.]{
Brandt A. L. Gaches,$^{1,2}$\thanks{E-mail: gaches@ph1.uni-koeln.de (BALG)}
Stefanie Walch,$^{1,3}$
Richard W\"{u}nsch$^{4}$
and Jonathan Mackey$^{5}$
\\
$^{1}$I. Physikalisches Institut, Universit\"{a}t zu K\"{o}ln, Z\"{u}lpicher Stra{\ss}e 77, 50937, K\"{o}ln, Germany\\
$^{2}$Department of Space, Earth and Environment, Chalmers University of Technology, Gothenburg SE-412 96, Sweden\\
$^{3}$Center for Data and Simulation Science (CDS), University of Cologne, www.cds.uni-koeln.de, Germany\\
$^{4}$Astronomical Institute of the Czech Academy of Sciences, Bo\v{c}n\'{i} II 1401, 141 00 Prague, Czech Republic\\
$^{5}$Centre for AstroParticle Physics and Astrophysics, DIAS Dunsink Observatory, Dunsink Lane, Dublin 15, Ireland
}

\date{Accepted XXX. Received YYY; in original form ZZZ}

\pubyear{2022}

\begin{document}
\label{firstpage}
\pagerange{\pageref{firstpage}--\pageref{lastpage}}
\maketitle

\begin{abstract}
X-ray radiation, in particular radiation between 0.1 keV and 10 keV, is evident from both point-like sources, such as compact objects and T-Tauri young stellar objects, and extended emission from hot, cooling gas, such as in supernova remnants. The X-ray radiation is absorbed by nearby gas, providing a source of both heating and ionization. While protoplanetary chemistry models now often include X-ray emission from the central young stellar object, simulations of star-forming regions have yet to include X-ray emission coupled to the chemo-dynamical evolution of the gas. We present an extension of the {\sc TreeRay} reverse raytrace algorithm implemented in the {\sc Flash} magneto-hydrodynamic code which enables the inclusion of X-ray radiation from 0.1 keV $< E_{\gamma} <$ 100 keV, dubbed {\rm XrayTheSpot}. {\sc XrayTheSpot} allows for the use of an arbitrary number of bins, minimum and maximum energies, and both temperature-independent and temperature-dependent user-defined cross sections, along with the ability to include both point and extended diffuse emission and is coupled to the thermochemical evolution. We demonstrate the method with several multi-bin benchmarks testing the radiation transfer solution and coupling to the thermochemistry. Finally, we show two example star formation science cases for this module: X-ray emission from protostellar accretion irradiating an accretion disk and simulations of molecular clouds with active chemistry, radiation pressure, protostellar radiation feedback from infrared to X-ray radiation.
\end{abstract}




\section{Introduction}
Molecular gas is subjected to radiation across the electromagnetic spectrum. Hard radiation, such as X-ray and gamma-ray radiation, can penetrate deep into molecular gas and drive the thermochemistry of dense gas \citep{Spitzer1968, maloney1996, yan1997, wolfire2022}. X-rays provide an important source of ionization in dense gas, driving the ion-neutral chemistry and providing heating through photo-electrons \citep{Lepp1983a, maloney1996, dalgarno1999}. Using the typical molecular gas photoabsorption cross sections \citep{maloney1996}, the $\tau = 1$ surface for 1 keV photons is approximately $4\times10^{21}$ cm$^{-2}$ (compared to $\approx 10^{-18}$ cm$^{-2}$ for UV radiation). However, the photoabsorption cross sections scale roughly as $E^{-2.5}$ \citep{mackey2019}, so harder radiation penetrates much further into the cloud. Therefore, in regions near bright X-ray sources, the cloud structure can become dominated throughout by the X-ray radiation.  Regions in which the thermochemistry is regulated primarily through X-ray radiation are often denoted as X-ray Dominated Regions (XDRs) \citep[coined by][]{maloney1996}, in analogue to photo-dissociation regions (PDRs).

X-ray radiation drives ionization primarily through secondary, induced processes. While the primary ionization cross sections are low, the resulting ejected fast electrons can produce a cascade of secondary ionizations and pumped far ultraviolet (FUV) radiation through the excitation and subsequent de-exictation of H and H$_2$ \citep{prasad1983, dalgarno1999, meijerink2005}. These fast electrons can also provide heating through photoelectric heating of dust grains. In this way, X-ray radiation acts in a very similar manner as cosmic rays, and untangling the two contributions can be difficult \citep[see e.g.][]{Meijerink2006}. However, due to the significantly larger interaction cross sections (and the strong energy and temperature dependence), the effect of X-ray radiation on the heating and ionization fraction are still noticeably different \citep{Meijerink2006}. There have been a plethora of investigations on the impact of cosmic rays on interstellar gas \citep[e.g.][]{Dalgarno2006, Ceccarelli2011, Indriolo2013, Bisbas2015, Gaches2019, Bisbas2023} showing that they play a crucial role in the chemistry of the interstellar medium. Due to the ubiquitiy of the inclusion of cosmic rays, it is relatively common to treat the chemical impact of X-rays through enhancing the cosmic ray ionization rate, although some studies separate the heating treatments \citep[e.g.][]{Harada2010, Viti2014, Walsh2015, Viti2017, Wang2021}.

Molecular clouds are immersed in a bath of X-ray radiation, with contributions from both external and internal sources. Externally, molecular gas can be irradiated through supernovae and their remnants \citep[e.g.][]{Yamane2018, Brose2022}, X-ray binaries \citep[e.g.][]{White1988, Remillard2006, Reig2011, mineo2012, lutovinov2013, giacobbo2018}, nearby activate galactic nuclei (AGN) \citep[e.g.][]{Sunyaev1993, Sunyaev1998, Harada2013, Churazov2017, Mingozzi2018, Cruz-Gonzalez2020}. Internally, young stellar objects, including embedded accreting protostars and more evolved T-Tauri stars \citep{calvet1998, Feigelson1999, Feigelson2007}, and high-mass stars just reaching the main sequence can become X-ray bright \citep[e.g.][]{Cassinelli1994}, whether through accretion or magnetic powered radiation or coronal emission. Finally, gas heated through feedback processes, such as winds and supernovae, can become warm enough to emit X-ray radiation while they cool \citep{raymond1977}. Observational X-ray surveys of molecular gas and star-forming regions show substantial amounts of diffuse emission and a sizable number of point sources \citep{Sunyaev1993, Feigelson2013, Townsley2014, Townsley2019}.

Despite their potential importance, their inclusion into simulations of molecular clouds has been sparse. There has been substantial focus on thermochemical models of protoplanetary disks \citep[e.g.][]{Glassgold1997, Igea1999, Ercolano2008a, Ercolano2009, Owen2011, Meijerink2012, Cleeves2017, Picogna2019, Waggoner2019} and models of molecular gas near external sources or compact objects \citep[e.g.][]{Krolik1983, Lepp1983b, Draine1991, Garcia-Burillo2010, Hocuk2010, Meijerink2011, Odaka2011, Orlando2011, mackey2019}. These methods typically utilize Monte Carlo methods \citep{Ercolano2008a, Odaka2011, Molaro2016, Walls2016, Cleeves2017}, or ray-trace schemes and focus primarily either on the inclusion of point sources or external radiation fields \citep[e.g.][]{Wise2011, mackey2019, Khabibullin2020}

In this paper, we will present an X-ray extension of the reverse ray tracing scheme {\sc TreeRay} \citep{wunsch2021}, which allows for the inclusion of an arbitrary number of point sources and diffuse radiation. The module is a {\sc TreeRay} extension of the diffuse X-ray module presented in \citet{mackey2019}, which enabled diffuse X-ray irradiation at the domain boundary. Our implementation enables up to 100 energy bins at arbitrary locations between 0.1 and 100 keV and temperature-dependent photoabsorption cross sections. In Section \ref{sec:methods} we give an overview of the X-ray {\sc TreeRay} algorithm, called {\sc XRayTheSpot}, and the coupling of it to the X-ray-driven chemistry. In Section \ref{sec:tests} we show the performance of the module with different radiation transfer tests and a benchmark against the {\sc Cloudy} code. In Sections \ref{sec:disk} and \ref{sec:mcsim} we demonstrate the use of this module for protostellar emission irradiating a surrounding disk and in a star formation simulation, respectively. Finally, in Section \ref{sec:discuss} we discuss the future extensions and scientific applications of {\sc XRayTheSpot}.

\section{Methods}\label{sec:methods}
Our new {\sc XRayTheSpot} module is able to treat radiation from 0.1~keV to 100 keV. We describe below in detail the adopted photoabsorption cross sections and the module's implementation within {\sc TreeRay}. 
\subsection{XRay Cross Sections}\label{sec:xsecs}
X-ray radiation is attenuated as it propagates through gas via a combination of photoionization, at lower energies, and the Compton process, at high energies. The previous module, described in \citet{mackey2019}, used the low-energy approximation for the X-ray cross section, $\sigma_x$, from \citet{panoglou2012}:
\begin{equation}\label{eq:analyxsec}
    \sigma_x = 2.27\times10^{-22} E^{-2.485}_{\gamma} \,\,{\rm cm^2}
\end{equation}
per H-nucleus, where $E_{\gamma}$ is the photon energy. However, this cross section is valid only for cold, neutral gas and for solar metallicity. We include now, as input during run time, temperature-dependent cross sections, which can be re-computed for problems with different metallicities. Decreasing the metallicity will primarily impact X-rays between 300 eV $\le E \le$ 10 keV due to the decrease in contributions from various metals. A benchmark comparison for different metallicities are presented below. For photoionization, we use the analytic fits from \citet{verner1995}, $\sigma_{\rm pi}$, where
\begin{equation}
    \sigma_{\rm pi} = \sigma_0 F(E_\gamma/E_0),
\end{equation}
where
\begin{equation}
    F(y) = \left [ \left ( y - 1\right )^2 + y_w^2 \right] y^{-Q} \left ( 1 + \sqrt{y/y_a}\right )^{-P},
\end{equation}
$y = E_\gamma/E_0$, $Q = 5.5 + l - 0.5$P, $l = 0,1,2$ is the subshell orbital quantum number, and $\sigma_0$, $E_0$, $y_w$, $y_a$ and P are fit parameters from the associated public ViZieR catalog\footnote{\url{https://cdsarc.cds.unistra.fr/viz-bin/cat/J/A+AS/109/125}}. However, to utilize these cross sections, the ionization level populations must be known. We assume collisional ionization equilibrium and use the {\sc ChiantiPy} package \citep{Dere2013}, using version 9 of the {\sc Chianti} atomic database \citep{dere1997, Dere2019} to compute the ionization fraction as a function of temperature. 

\begin{figure*}
    \centering
    \includegraphics[width=0.95\textwidth]{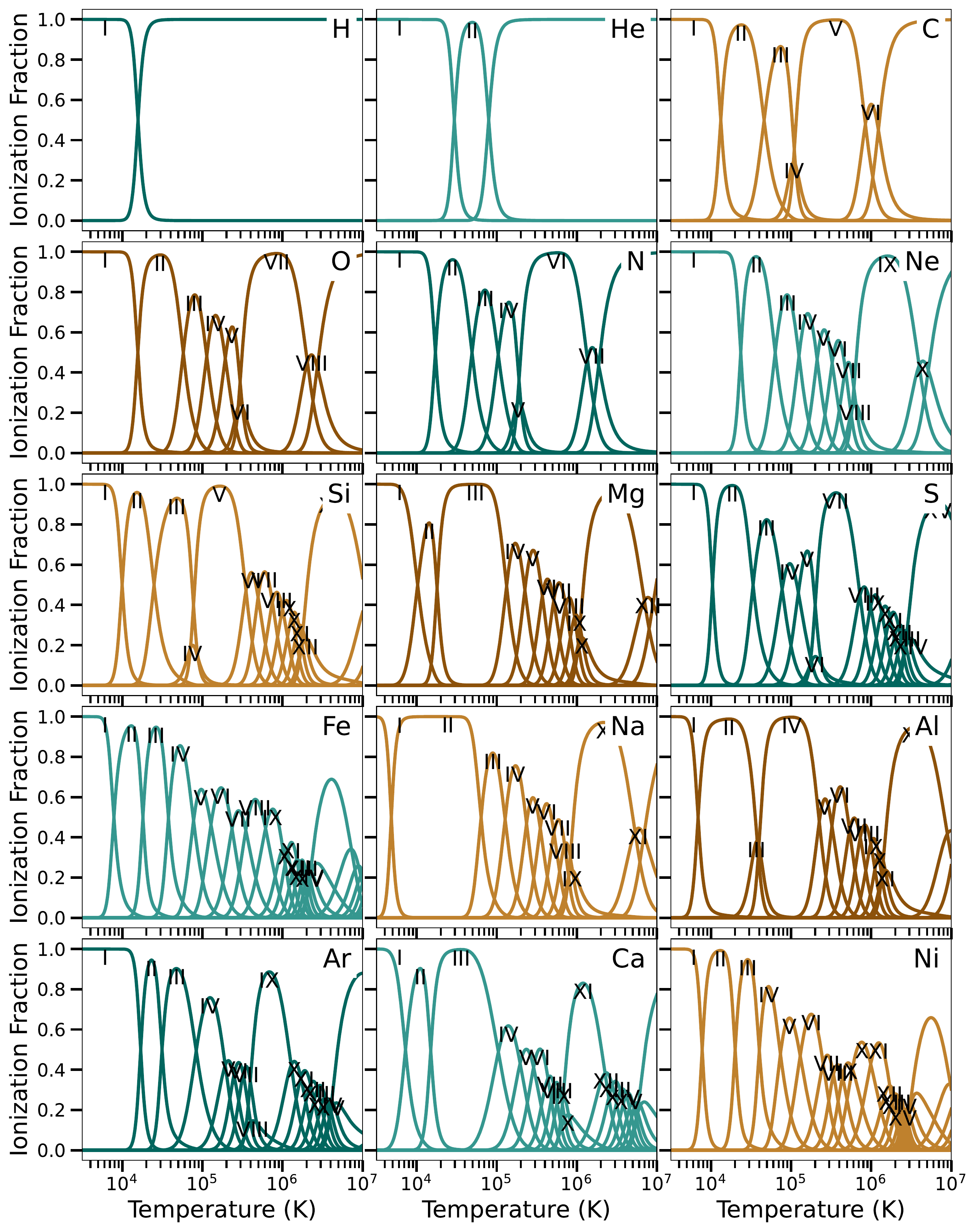}
    \caption{\label{fig:ionbalance}Ionization fraction for different elements as a function of temperature. Annotated in the text are the peaks of different ionization levels for each element.}
\end{figure*}

Figure \ref{fig:ionbalance} shows the equilibrium ionization fractions as a function of temperature for the 15 different elements (see Table \ref{tab:elem}) we include in the cross sections. These computations show that, particularly for $T > 10^5$ K, there are multiple ionization states for metals which contribute to the photoionization cross section. 

We also include the cross section for the Compton effect, which becomes particularly important at higher energies. We use the total Klein-Nishina (KN) cross section \citep{klein1929, longair2011}, $\sigma_{\rm KN}$,
\begin{equation}
    \sigma_{\rm KN} = \pi r_e^2 x^{-1} \left \{   \left [ 1 - \frac{2 (x + 1)}{x^2} \right ] \ln (2x+1) + \frac{1}{2} + \frac{4}{x} - \frac{1}{2(2x+1)^2} \right \},
\end{equation}
where $r_e$ is the classical electron radius and $x = E_\gamma/(m_e c^2)$. While most applications will be in the limit of Thomson scattering, we include the full Compton cross sections to enable more flexibility in the choice of energy bins. In the cross-section plots, we show the total Compton cross-section weighted by the number of free electrons contributed by each species. The total cross section, $\sigma_x$, is thus
\begin{equation}
    \sigma_x(E) = \sum_i^{N_{\rm elem}} x_i \sigma_{\rm pi,i}(E) + \sigma_{\rm KN,i}(E),
\end{equation}
where $x_i$ is the abundance of element $i$ with respect to hydrogen and the sum is carried out including the cross sections for the $N_{\rm elem}$ elements. Table \ref{tab:elem} shows the elements we include in the photoabsorption cross section and their fiducial abundances relative to hydrogen.

\begin{figure}
    \centering
    \includegraphics[width=0.5\textwidth]{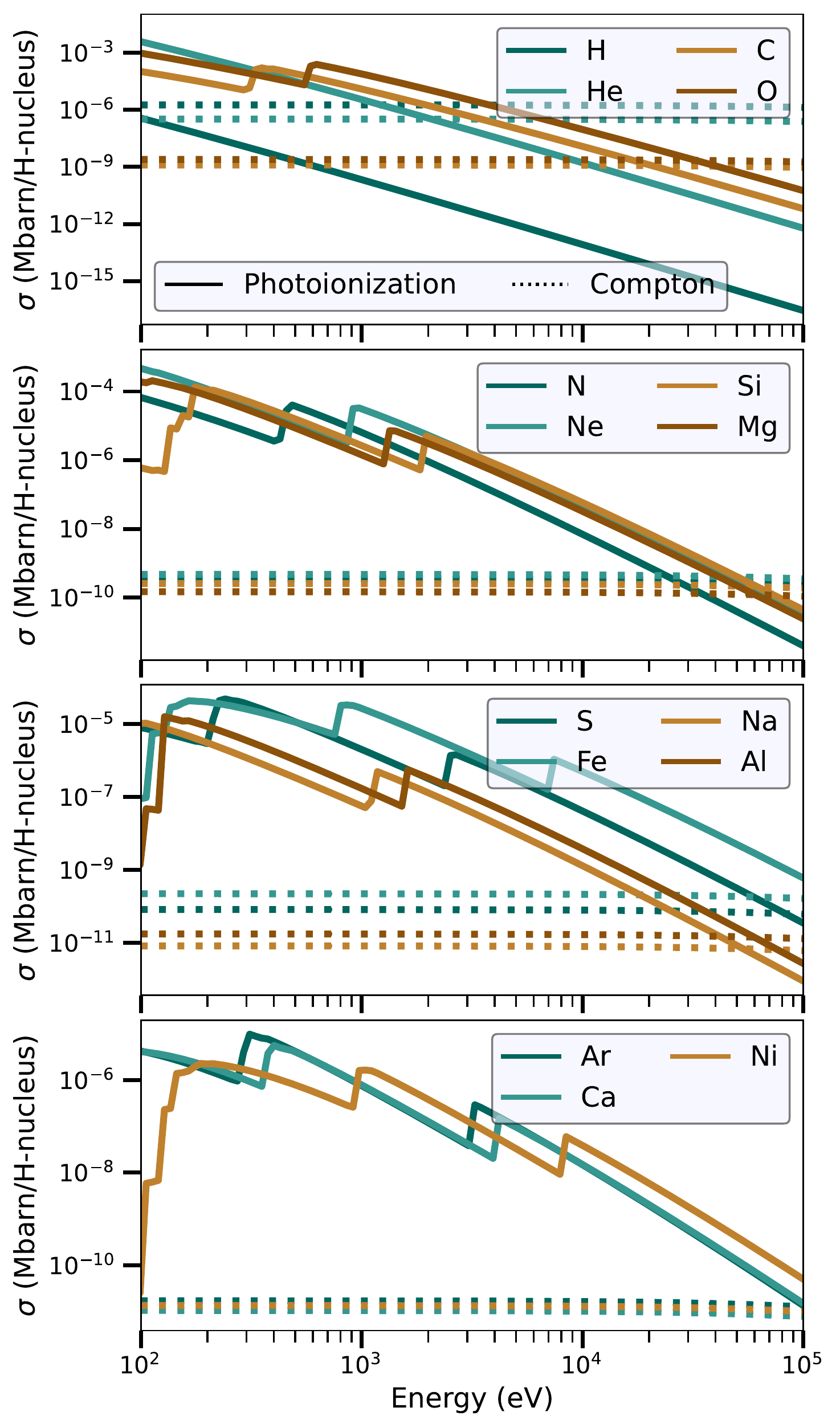}
    \caption{\label{fig:xsecspec}Photoionization (solid) and Compton process (dashed) cross sections for each element as a function of energy, assuming thermal ionization equilibrium at $T = 10^5$ K. Each elemental contribution is weighted by the assumed abundance with respect to hydrogen.}
\end{figure}

\begin{figure}
    \centering
    \includegraphics[width=0.5\textwidth]{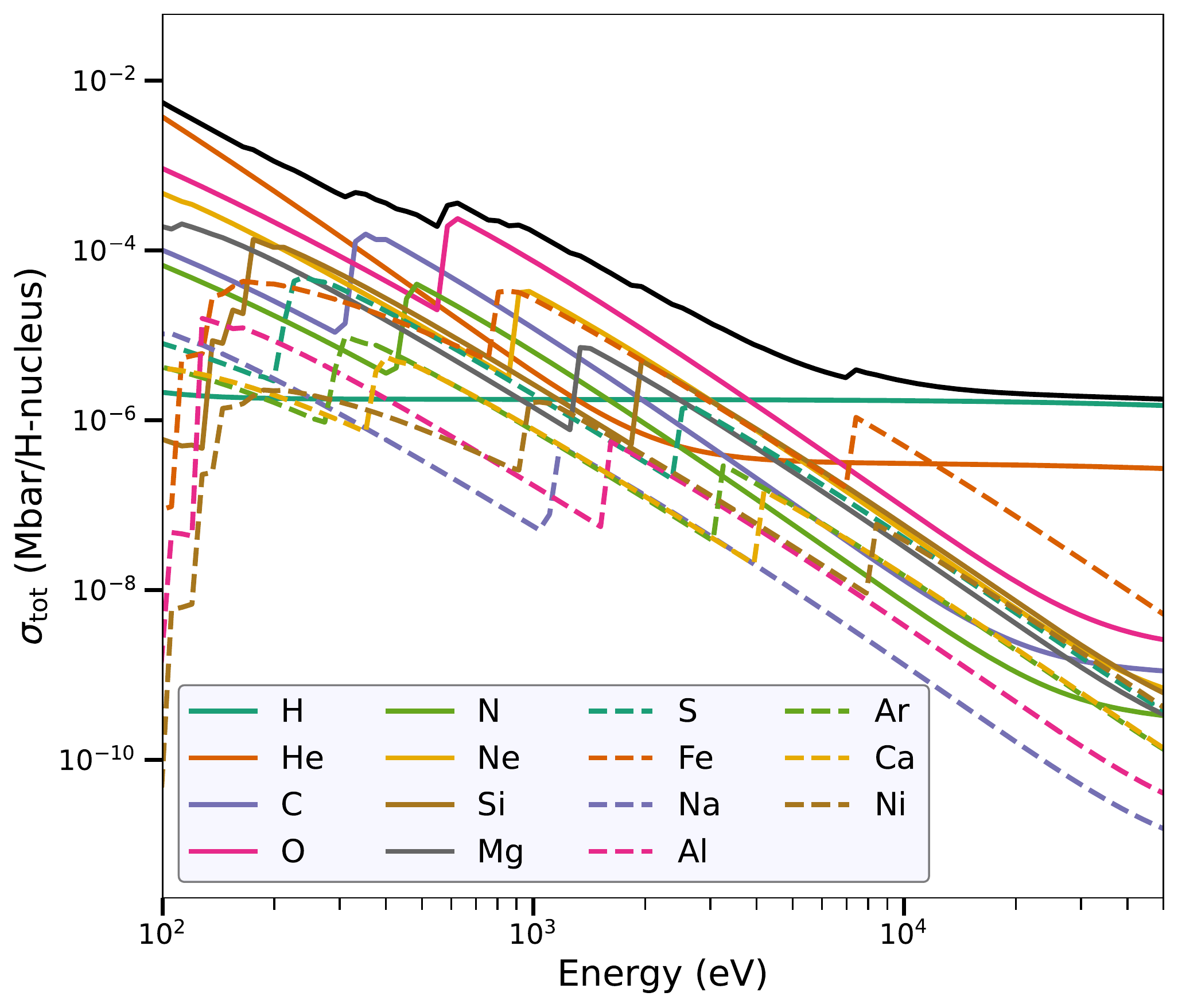}
    \caption{\label{fig:totalcross}Total photo-absorption cross section (black) with each element contribution highlighted (colors) as a function of energy for gas at temperature, $T = 10^5$ K. Each elemental contribution is weighted by the assumed abundance with respect to hydrogen.}
\end{figure}

Figure \ref{fig:xsecspec} shows the photoionization cross sections and free-electron contributions to the Compton cross section as a function of energy for gas with $T = 10^5$ K. As shown, for some elements, the Compton cross section becomes more important than photoionization, in particular for hydrogen and helium above 1 keV, and for carbon and oxygen above 30 keV. For hydrogen, the Compton effect is dominant due to the negligible neutral fraction at T = $10^5$ K. Figure \ref{fig:totalcross} shows the total cross section as a function of energy for $T = 10^5$ K and each of the total elemental contributions. Here, the X-ray photoabsorption cross section is dominated by helium ($<0.3$ keV), then carbon ($0.3 - 0.6$ keV) and oxygen ($0.8 - 4$ keV). At energies above 4 keV, the hydrogen and helium Compton cross sections dominate with a contribution from the iron photoionization cross section around 10 keV. However, many of the metals contribute equally to the total cross section around 1 keV. 

\begin{figure}
    \centering
    \includegraphics[width=0.5\textwidth]{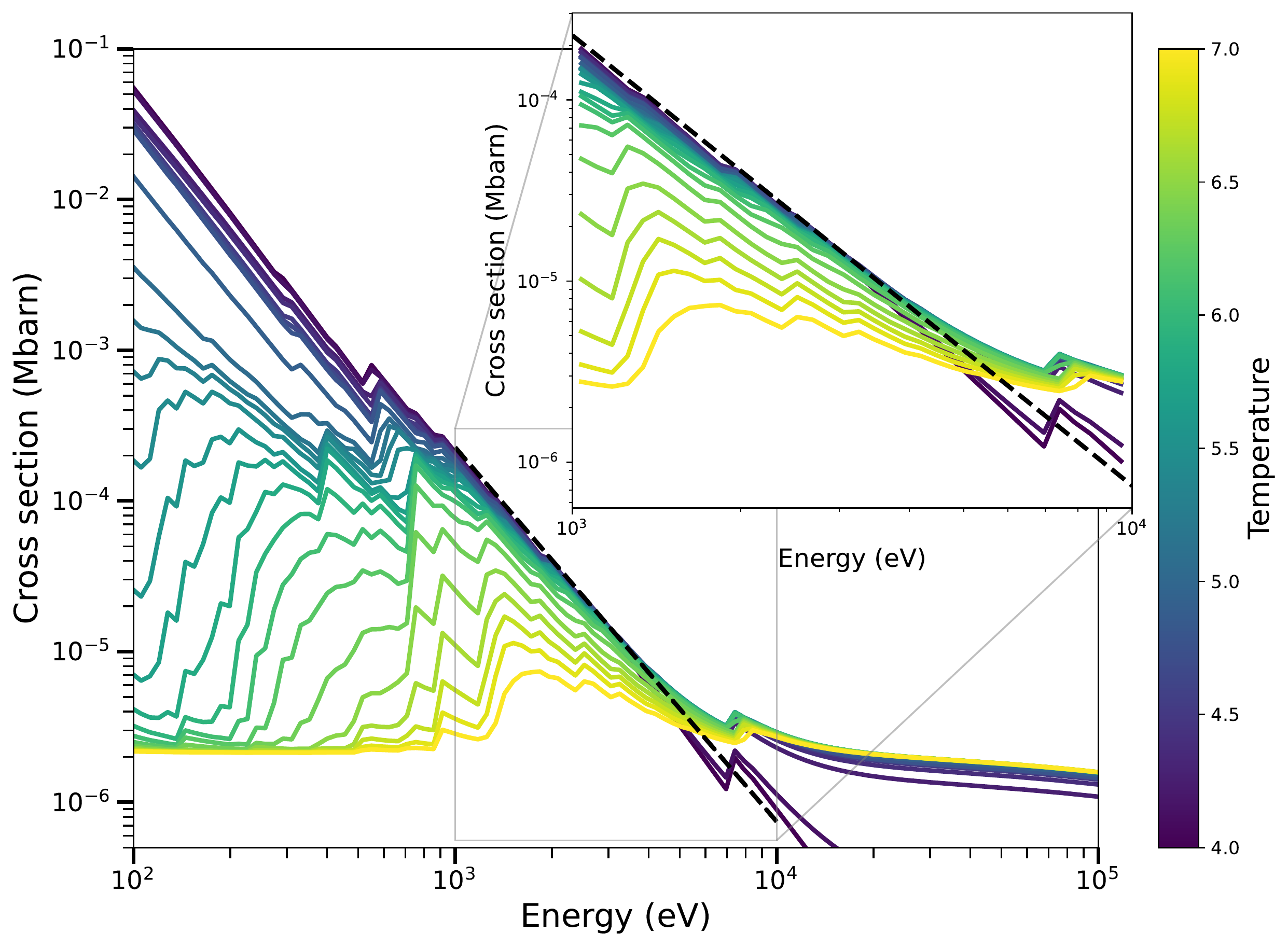}
    \caption{\label{fig:xsectemp}Total photoabsorption cross section as a function of energy and temperature (color). The black dashed line shows the previously used analytic cross section from \citet{panoglou2012}. Inset: Zoom-in to 1 -- 10 keV.}
\end{figure}

Finally, Figure \ref{fig:xsectemp} shows $\sigma_x$ as a function of energy and temperature from $T = 10^4$ to $10^7$ K. At low temperatures, we recover the analytic cross section previously used, although around 10 keV there is an increase in the cross section due to iron. However, the rather significant temperature dependence highlights the necessity of including a temperature dependent cross section: \emph{at high temperatures, the higher thermal ionization state leads to a reduction of nearly two orders of magnitude in the cross section}, thereby making the gas significantly more optically thin to the X-ray radiation, producing less heating and enabling more X-rays to escape. Below $10^4$ K, the cross section does not noticeably change since hydrogen is not significantly collisional ionized. Therefore, for the results of this paper, for colder gas, we use the $T = 10^4$ K photo-absorption cross section. The module though allows for the user to define their own temperature dependent cross sections across any temperature range.

For a given set of energy bins, $\{(E_{l,i}, E_{r,i})\}$, where $i = 1, N_{\rm bin}$ for $N_{\rm bin}$ bins, we define:
\begin{equation}
    E_{c,i} = \frac{1}{2} \left ( E_{l,i} + E_{r,i} \right ),
\end{equation}
where $E_{l,i}$ is the left bound of the $i^{\rm th}$ bin, $E_{r,i}$ is the right bound, and $E_{c,i}$ is the midpoint of the bin. We derive bin-averaged cross sections, such that
\begin{equation}
    \exp \left ( -\frac{\langle \sigma_{X,i} \rangle }{\sigma_c} \right ) = \frac{1}{E_{l,i} - E_{r,i}} \int_{E_{l,i}}^{E_{r,i}} \exp \left ( -\frac{\sigma_x(E_\gamma)}{\sigma_c} \right ) dE,
\end{equation}
where $\langle \sigma_{x,i} \rangle$ is the bin-averaged cross section for bin, $i$, and $\sigma_c = \sigma_x(E_{c,i})$. Our fiducial tests use $N_{\rm bin} = 8$ between 1 -- 10 keV using logarithmically spaced bins.

All of these cross section data, and the initialization and storage of the bins and bin-averaged cross sections are kept in a new {\sc Flash module}, {\sc XrayCommon}. {\sc Flash} is a highly module public magneto-hydrodynamic code \citep{fryxell2000} written in {\sc Fortran} and highly-scalable with {\sc MPI}. The scripts necessary to compute the X-ray cross sections are publicly available on {\sc GitHub}\footnote{\faGithubSquare \, \url{https://github.com/AstroBrandt/XRayCrossSections}}. This module enables the coupling of X-ray physics to multiple other modules. Plasma models and the necessary X-ray data are also stored within this module, as a unified location.

\begin{table}
    \caption{Elements included in our photoabsorption cross section calculation and their fiducial abundances, $A_X$, reported as $A_X = \log (N_x/N_H) + 12$ \citep{Asplund2009}.}
    \label{tab:elem}
    \centering
    \begin{tabular}{c|c}
        Element & Abundance ($A_X$) \\
        \hline
        H & 12 \\
        He & 10.986 \\
        C & 8.443 \\
        O & 8.783 \\
        N & 7.913 \\
        Ne & 8.103 \\
        Na & 6.353 \\
        Mg & 7.593 \\
        Al & 6.523 \\
        Si & 7.573 \\
        S & 7.193 \\
        Ar & 6.553 \\
        Ca & 6.383 \\
        Fe & 7.503 \\
        Ni & 6.283 
    \end{tabular}
\end{table}

\subsection{TreeRay}\label{sec:treeray}
{\sc TreeRay} is a novel reverse ray tracing scheme, described fully in \citet{wunsch2021}, implemented in {\sc Flash}. Simply, {\sc TreeRay} enables an efficient method to compute the contributions of radiation from every cell, for every cell. It does so through the combination of a reverse ray-trace algorithm with a tree \citep{wunsch2018}, which also currently is used in the gravity solver. Below we describe briefly the different aspects of the {\sc TreeRay} algorithm and {\sc XRayTheSpot} extension and refer the reader to \citet{wunsch2021} for more details.

\subsubsection{Building the Tree}
The foundation of the {\sc TreeRay} algorithm is an octtree which stores all necessary variables for the various {\sc TreeRay} modules. At minimum, the tree stores the mass and center of mass coordinates for the respective cell, or leaf, or higher nodes. For {\sc XRayTheSpot}, two further quantities are stored onto the tree: the bin-integrated X-ray luminosity in each energy bin and the gas temperature. While the bin-integrated X-ray luminosity is purely additive, the temperature is stored as a mass-weighted average of each set of eight sub-nodes (or leaves).

\subsubsection{Ray Structure}
Before the tree walk is executed for a given cell, rays are generated by casting $N_{\rm pix} = 12 N_{\rm side}^2$ rays from each cell using directions defined by the {\sc HealPix} \citep{gorski2005} algorithm. {\sc HealPix} tessellates the unit sphere into areas representing equal solid angles with a unit vector pointing to the center of each of these surface areas from the sphere's center. {\sc TreeRay} allows for $N_{\rm side} = 1, 2, 4, 8,...$, with higher values representing higher angular resolution. The rays are split into $N_r$ evaluation points, set by the grid resolution, $\Delta x$, the allowed length of the ray, $L_{\rm ray}$, which is set to three-dimensional diagonal of the computational domain, and a free parameter, $\eta_R$. 

Along each ray, the radial coordinate point of the i$^{\rm th}$ evaluation point is
\begin{equation}
    r_i = \frac{\Delta x i^2}{2 \eta_R^2},
\end{equation}
leading to segments with increasing lengths. This behavior coincides well with the geometric acceptance criterion described below for deciding whether or not to accept a tree node. The total number of evaluation points is
\begin{equation}
    N_R = \eta_R \times \text{floor} \left ( \sqrt{\frac{2 L_{\rm ray}}{\Delta X}}\right ) +1.
\end{equation}

\subsubsection{Tree Walk}
The mapping of the cells/nodes onto the rays requires two factors: a multipole acceptance criterion (MAC) and a weighting function to map from the tree onto the different radial evaluation points. When the MAC is met, a node is accepted and used. The simplest MAC is the Barnes-Hut (BH) geometric MAC \citep{barnes1986}, where a node of size $h_n$, at a distance $d$, from the cell is opened if 
\begin{equation}
    h_n/d < \theta_{\rm lim}
\end{equation}
where $\theta_{\rm lim}$ is a user-defined opening angle with a sensible choice being $\theta_{\rm lim} = \sqrt{4\pi/N_{\rm pix}}$\footnote{The resulting $\theta_{\rm lim}$ for $N_{\rm side} = 1, 2, 4, 8$ is $1.0, 0.5, 0.25, 0.125$, respectively. For the results of this paper, we adopted these recommended values of $\theta_{\rm lim}$ for the corresponding $N_{\rm side}$. See \citet{wunsch2018} for $\theta_{\rm lim}$ resolution tests in the context of {\sc TreeRay/OpticalDepth}}. We also utilize the `Src MAC' \citep{wunsch2021}, where a node with sources is opened if
\begin{equation}
    h_n/d < \theta_{\rm src},
\end{equation}
where $\theta_{\rm src}$ is a user-defined parameter. 

Quantities on the tree are mapped onto the radial evaluation points of a ray through the use of kernels. We utilize both a piece-wise third-order polynomial, $W_p(\delta)$, and a kernel derived to ensure it meets the requirements of the radiation transfer equation, $W_f(\delta)$, where $\delta = (r_i - d)/h_n$ and $d$ is the distance from the node center of mass and the ray evaluation point \citep[see][]{wunsch2021} and $h_n$ is the node's linear size. The node quantites are weighted by the overlap of the volume of the ray segment and the node.

Following the tree walk, the rays from a cell outward store the mass, center of mass, gas temperature and bin-integrated luminosities. These provide all the necessary information to solve the equation of radiation transfer along each ray.

\subsubsection{Solving the Radiation Transfer Equation}
Along the rays, the one-dimensional radiation transfer equation is solved:
\begin{equation}
    \frac{dI_\nu}{ds} = -\epsilon_\nu + \alpha_\nu I_\nu
\end{equation}
where $s$ is the distance along the ray, and $\epsilon_\nu$ and $\alpha_\nu$ are the emission and absorption coefficients. The band-integrated flux, $J(E)$, irradiating a cell $i$ due to the band-integrated luminosity, $L_X$, emitting from node $j$ can be simply written
\begin{equation}
    J_{ji}(E) = L_{X, j}(E)\frac{e^{-\tau_x}}{4\pi r_{ij}^2}
\end{equation}
where $r_{ij}$ is the distance between the centers of cell $i$ to node $j$ and
\begin{equation}
    \tau_x = \int n_{\rm H}(s)\sigma_x(E, T(s)) ds \approx \sum_k \frac{\rho_k}{\mu m_H} \left <\sigma_X(E, T_k) \right > \delta s
\end{equation}
is the X-ray opacity between cell $i$ and node $j$ and $\rho_k$ is the density at evaluation point $k$ along the ray, $n_{\rm H}$ is the hydrogen nuclei density, and $\delta s = r_k - r_{k - 1}$. We store the solution as an energy density, $\varepsilon_x = J/c$, where $c$ is the speed of light, onto the grid to be used in chemistry, described below. The total energy density is the sum over the {\sc HealPix} rays:
\begin{equation}
    \varepsilon_i = \sum_k^{N_{\rm pix}} \varepsilon_{ki}(E).
\end{equation}
For the solution, we also store the cell mass and temperature onto the tree and map these to the rays using $W_p$. In our algorithm, no assumption is made with respect to what produces the X-ray energy density, enabling both point sources (with their radiation spread over their host cells) and diffuse emission produced via cooling of hot gas in the cell. 

\subsection{Pre-existing Chemistry}\label{sec:chem}
We briefly describe here the previous treatment of X-ray radiation within the chemical network (see \citet{mackey2019} for more details). The chemical network consists of 17 species, of which 9 are solved numerically and the rest are followed through conservation equations. We solve the non-equilibrum species \ce{H+}, \ce{H2}, \ce{C+}, \ce{CO}, \ce{HCO+}, \ce{CHx}, \ce{OHx}, \ce{He+} and \ce{M+}. \ce{CHx} is a proxy species for simple hydrocarbons, e.g. \ce{CH}, \ce{CH2}, etc, and simple ions \ce{CH+}, \ce{CH2+}, etc. Similarly, \ce{OHx} is a proxy for \ce{OH}, \ce{H2O} and ions \ce{OH+}, \ce{H2O+}, etc. \ce{M} is a proxy for metals that can become the primary source of electrons in shielded regions of molecular clouds, where reaction rates treat \ce{M} as \ce{Si}. The network is primarily based on the `NL99' network of \citet{glover2012}, which uses the hydrogen chemistry from \citet{glover2007a, glover2007b} with the CO chemistry of \citet{nelson1999} including updated reaction rates from \citet{gong2017}. For this work, all photodissociation rates have been updated using the KIDA astrochemistry database \citep{wakelam2012}.

X-ray radiation is coupled to the thermochemistry through the following primary processes \citep[see also][]{mackey2019}:
\begin{itemize}
    \item Dust heating, following the analytic prescription in \citet{yan1997}.
    \item Primary ionization of a species by X-rays. Note though that is is relatively unimportant for our considered species, and plays a minor role in the heating and ionization for hydrogen species and helium.
    \item Secondary ionization through collisional ionization by fast electrons produced following primary ionizations \citep[e.g.][]{dalgarno1999, meijerink2005}.
    \item Induced FUV radiation generated by \ce{H2}, which is collisionally excited by fast electrons and the subsequent ionizations and dissociations \citep{prasad1983, gredel1987, maloney1996, meijerink2005}.
    \item Coulomb heating of the gas via energy exchange between the produced fast electrons and other charged particles \citep{dalgarno1999}.
\end{itemize}
These processes have all been generalized for the arbitrary number of energy bins and the temperature-dependent cross sections. The input X-rays are computed by the {\sc XRayTheSpot} module. Since we use band-integrated radiative variables, the heating parameter for a particular cell $i$ due to the impinging X-ray radiation is
\begin{equation}
    H_{x,i} = \sum_n^{N_{\rm bin}} j_{i}(E_n) \left <\sigma_{x}(E_n, T_i) \right>.
\end{equation}

\section{Tests and Benchmarking}\label{sec:tests}
Here we show various numerical tests of the radiation transfer and a benchmark of the thermochemsitry against {\sc Cloudy}. For our benchmarks, we fiducially use 8 bins, logarithmically spaced between 1 - 10 keV, following \citet{meijerink2005}.
\subsection{Point Source Test}
Our first test is a single central point source with a constant luminosity distribution, $L_{x,n} = 1$ L$_{\odot}$ for all $N_{\rm bin}$ bins, embedded in a volume with a uniform density of $n(H) = 2\times10^3$ cm$^{-3}$ and a spatially constant temperature $T = 10$ Kelvin in a (30 pc)$^3$ volume. We use a constant luminosity to better compare the solutions of different energy bins. We then compute the radial profiles of the energy density and compare against the analytic solution:
\begin{equation}
    J(E, r) = L_x(E)\frac{e^{-\sigma_x(E) \rho r}}{4\pi r^2}
\end{equation}
where the energy density, $\epsilon = J(E)/c$. 

\begin{figure*}
    \centering
    \includegraphics[width=0.9\textwidth]{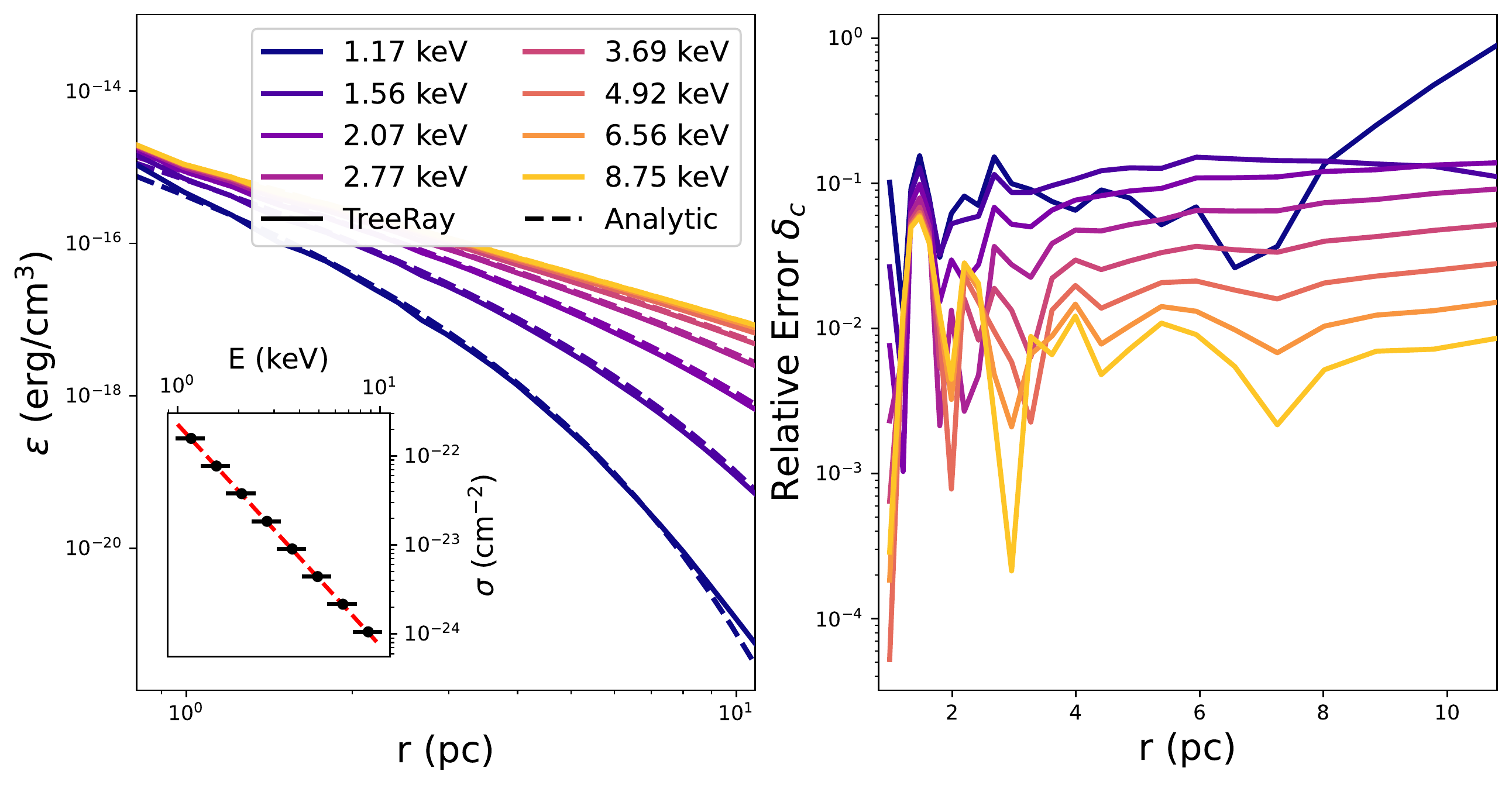} 
    \caption{\label{fig:singprof}Radial profile test for a single source in a constant density and temperature medium. Left: Radiation density versus radius for each bin for the {\sc TreeRay} (solid) and analytic solution (dotted). Left inset: Bin-averaged cross sections (black points) and the analytic cross section in Eq. \ref{eq:analyxsec}. Right: Relative errors for each bin as a function of radius.}
\end{figure*}

Figure \ref{fig:singprof} shows the performance of {\sc XRayTheSpot} for a single bright point source as a function of radius. The results in figure \ref{fig:singprof} used $\eta_R = 4$, $N_{\rm side} = 8$ and $N_{\rm block} = 8$, where $N_{\rm block}$ is the number of blocks of cells per spatial dimension, and one block consists of a cube of $8^3$ cells. The radial range was chosen that for the lowest energy bin, the emission transitions from optically thin to strongly optically thick, with the maximum radius corresponding to $\tau(E = 1.17 {\rm eV}) \approx 10$. The left panel shows the comparison between the ray trace solution and the analytic solution. These solutions agree well with each other with the lines largely overlapping. The right panel shows the relative error, defined as
\begin{equation}
\delta_c = \frac{|c\varepsilon- J(E, r)|}{J(E, r)}
\end{equation}
where $\varepsilon$ is the solution from {\sc XRayTheSpot}. The relative error is rather insensitive to the optical depth but more sensitive to how strongly the radiation field is coupled to the gas (e.g. the magnitude of the photoabsorption cross section). The 10\% error shown for the most optically thick bin at low energies is due to the mapping of the density structure onto the rays using the kernel. For X-ray optical depths greater than $\tau_x \approx 10$, the error starts to increase towards unity, but at these energy densities, the X-rays have a negligible impact on the thermochemistry. Therefore, these relative differences will have no discernible impact on the thermochemical evolution of the gas. 

\begin{figure}
    \centering
    \includegraphics[width=0.45\textwidth]{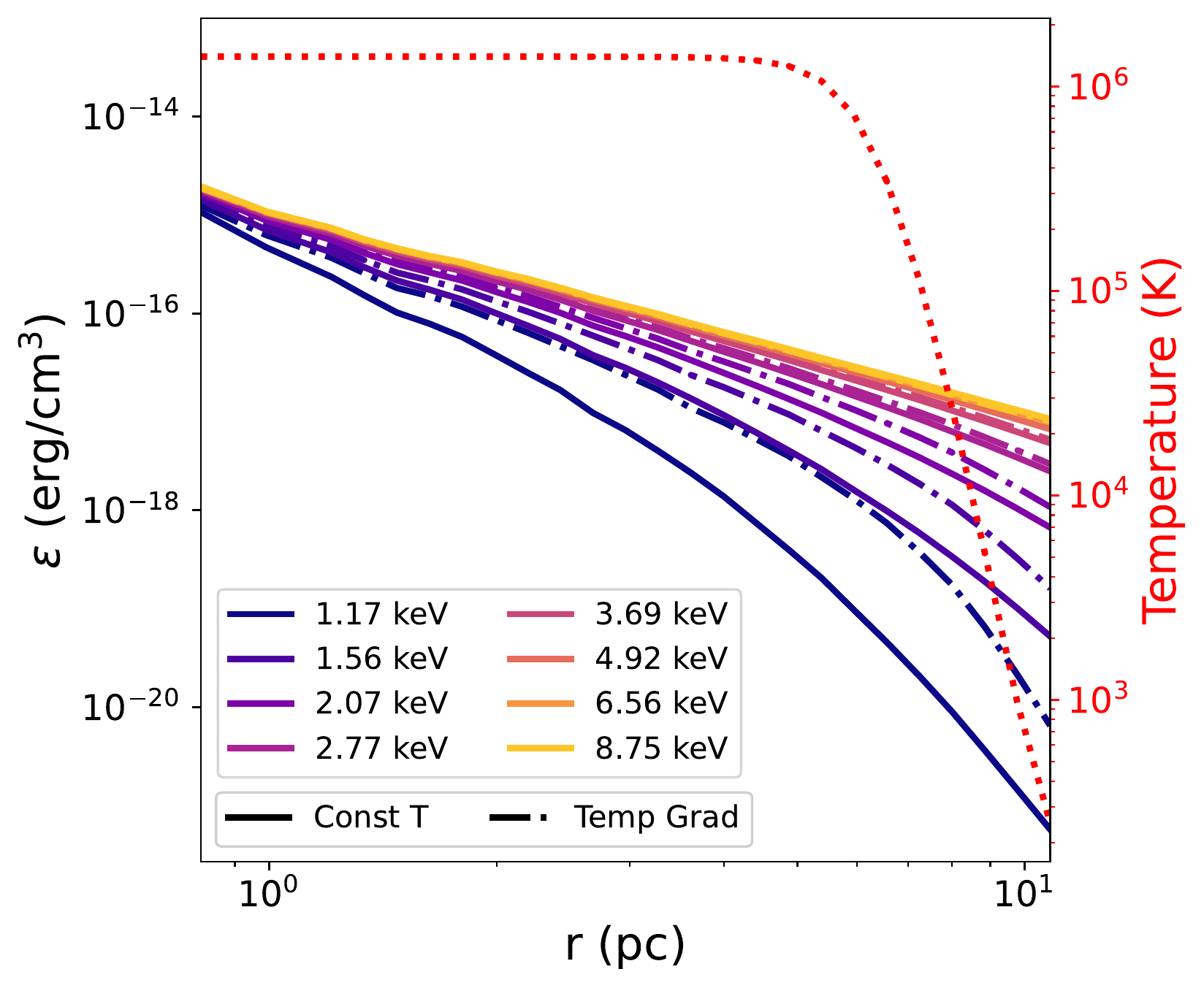} 
    \caption{\label{fig:tgradprof}X-ray energy density versus radius for single point source. The solid line uses the constant temperature at $T = 10$ K (same as Figure \ref{fig:singprof}) while the dashed-dotted line uses the temperature profile shown by the red dotted line.}
\end{figure}

In order to highlight the differences of the new module with the previous cross section implementation presented in \citet{mackey2019}, we perform a second calculation imposing a strong temperature gradient such that the radial temperature profile is
\begin{equation}
    T(r) = 5\times10^5 \left [ 1 - \tanh(r - 8 {\rm \, pc})\right ] + 100 \, {\rm K}.
\end{equation}
This temperature is  artificial and chosen such that the X-ray radiation transitions from optically thin to optically thick due to the change in cross section. Figure \ref{fig:tgradprof} shows the result of this comparison and as expected, the emission for the lower energy bins is up to an order of magnitude greater than the low-temperature solution and maintains an $r^{-2}$ trend until a much greater radius.

\begin{figure*}
    \centering
    \begin{tabular}{cc}
    \includegraphics[width=0.5\textwidth]{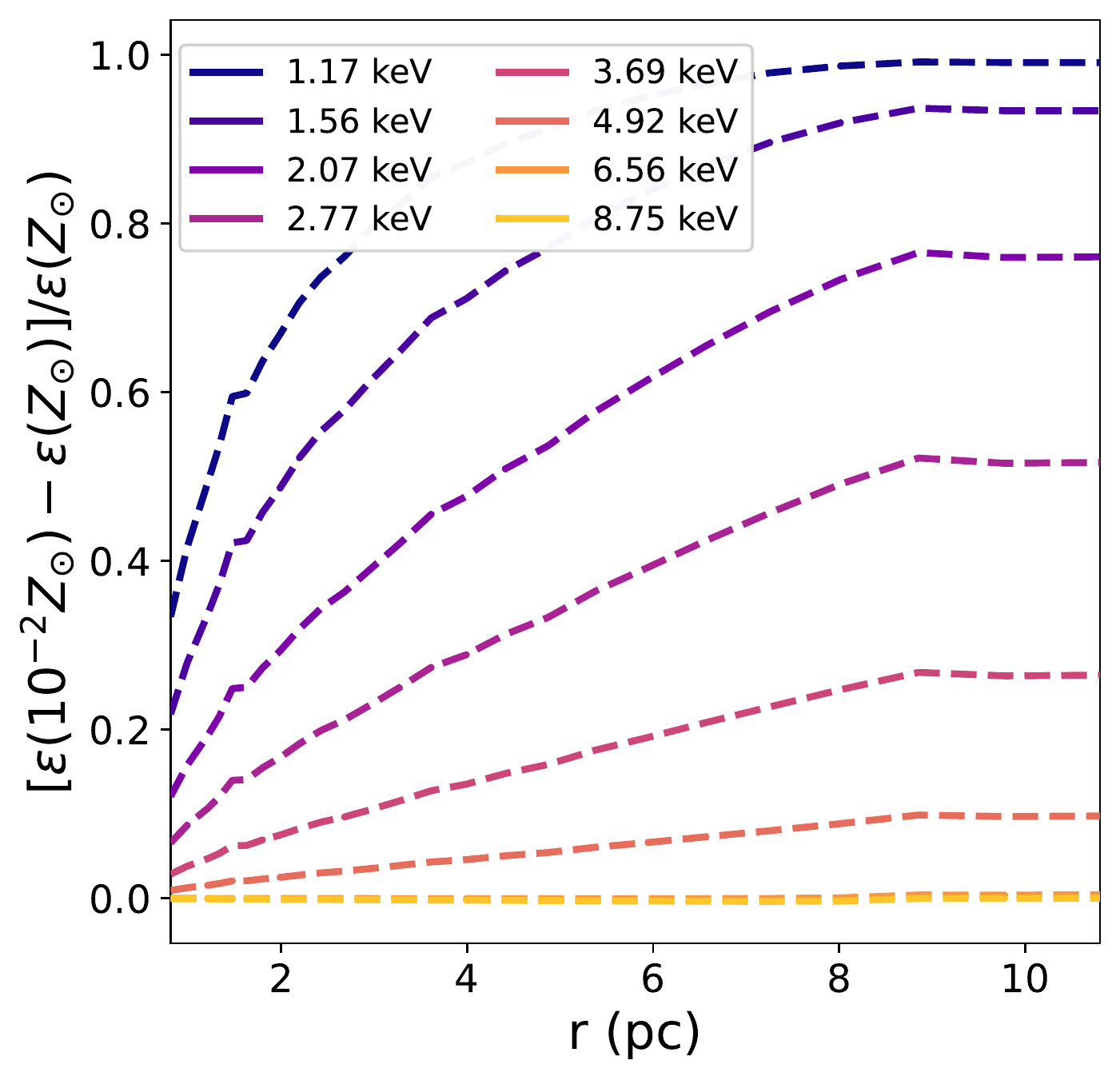} & 
    \includegraphics[width=0.5\textwidth]{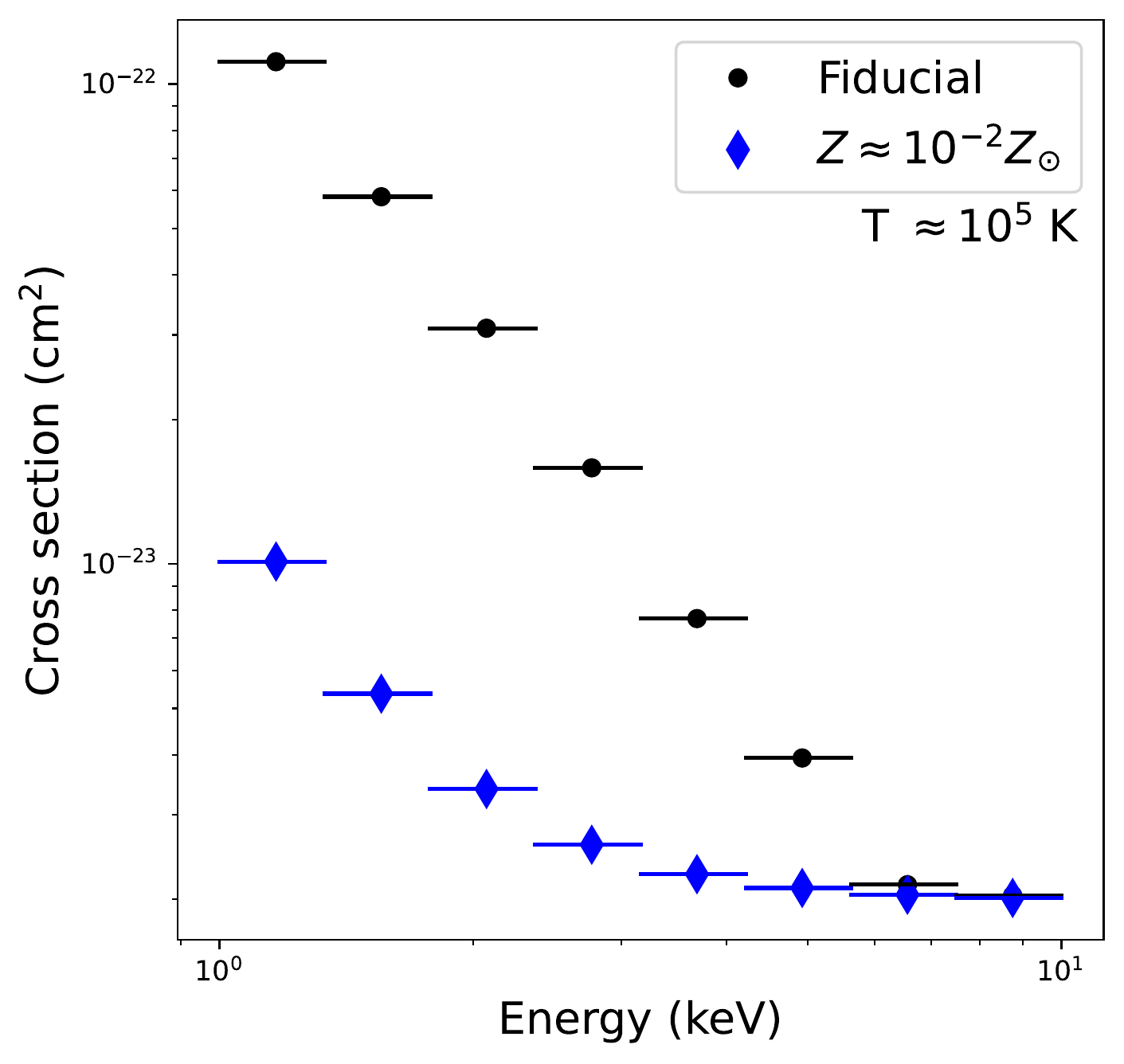}
    \end{tabular}
    \caption{\label{fig:metalDiff} Left: Relative deviations in the energy density between the fiducial abundances and a model with $Z = 10^{-2}Z_{\odot}$, defined as $[\varepsilon(10^{-2}Z_{\odot}) - \varepsilon(Z_{\odot})]/\varepsilon(Z_{\odot})$. Right: The cross sections at $T \approx 10^5$ K for the fiducial abundances (black) and $Z = 10^{-2}Z_{\odot}$ (blue).}
\end{figure*}

Fiducially, we assume the abundances shown in Table \ref{tab:elem}. We run an additional problem using the temperature profile above with a metallicity a factor of 100 lower. Figure \ref{fig:metalDiff} shows the relative differences in the energy densities for the two different metallicities and the cross sections at $T = 10^5$ K. The emission at very low metallicity is significantly enhanced due to the lack of photo-absorption by metals (see Fig. \ref{fig:totalcross}). For the highest energy bins, there is very little difference due to the cross section being by Compton scattering. 

\begin{figure*}
    \centering
    \begin{tabular}{cc}
        \includegraphics[width=0.45\textwidth]{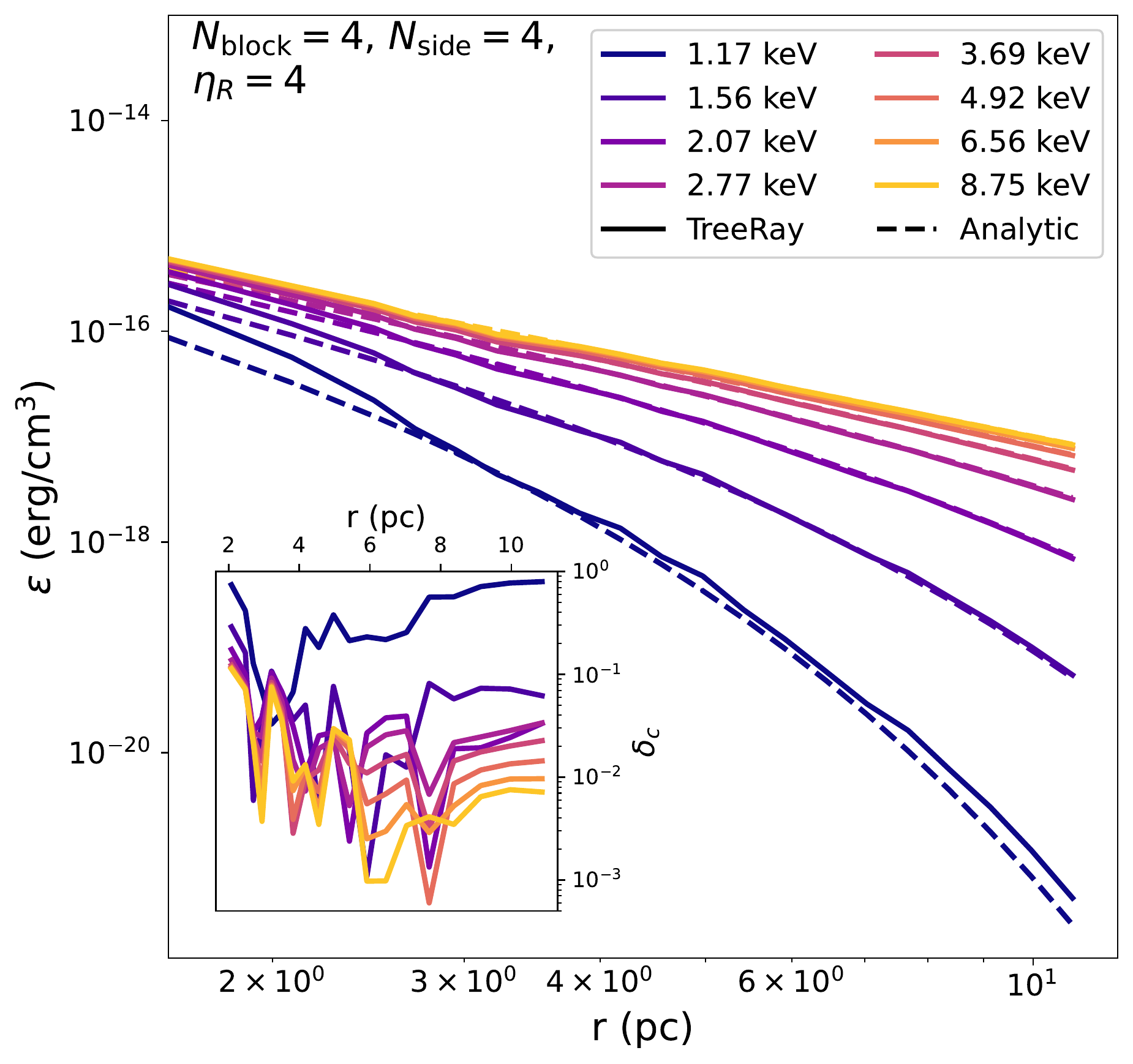} & \includegraphics[width=0.45\textwidth]{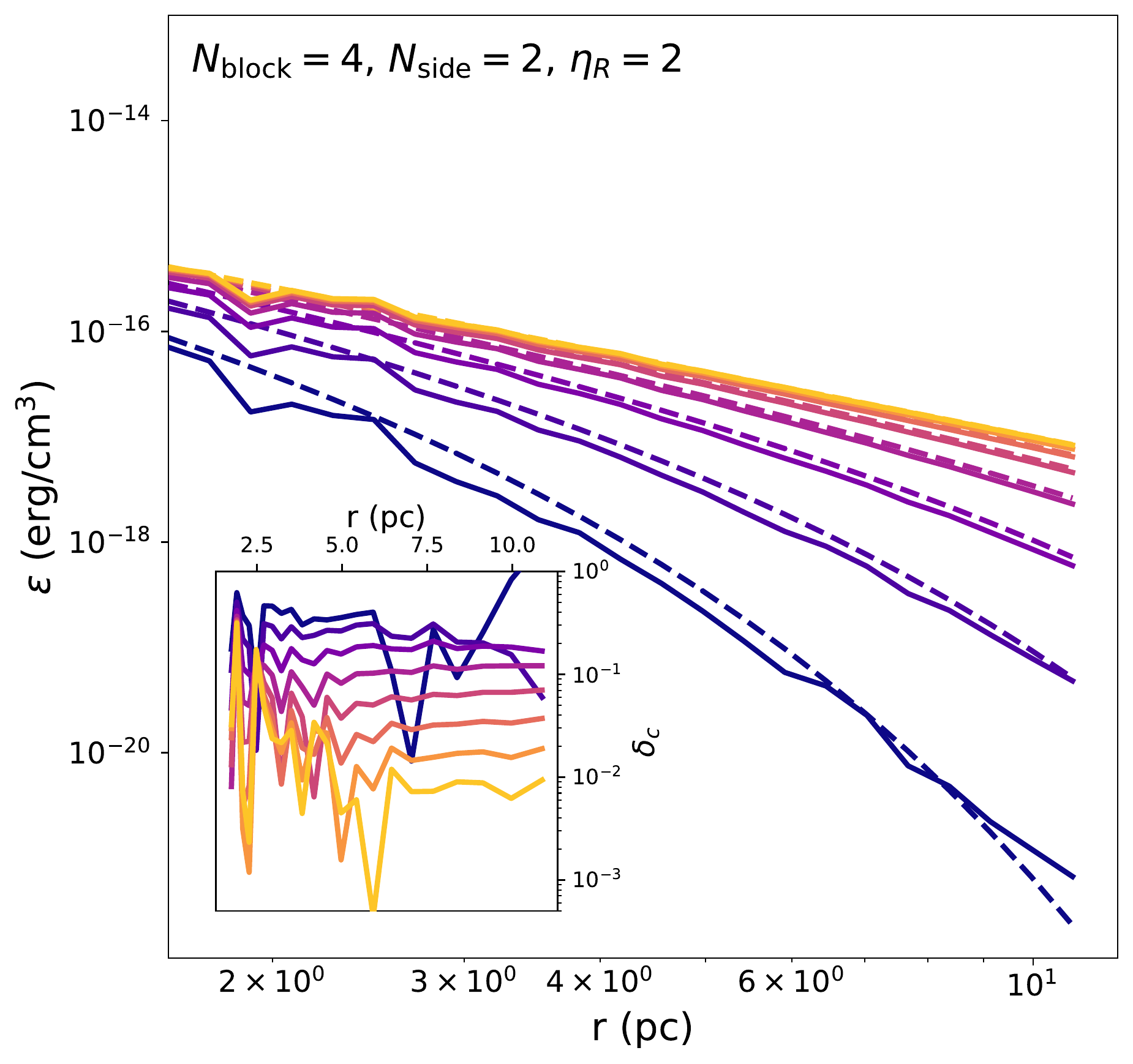} \\
        \includegraphics[width=0.45\textwidth]{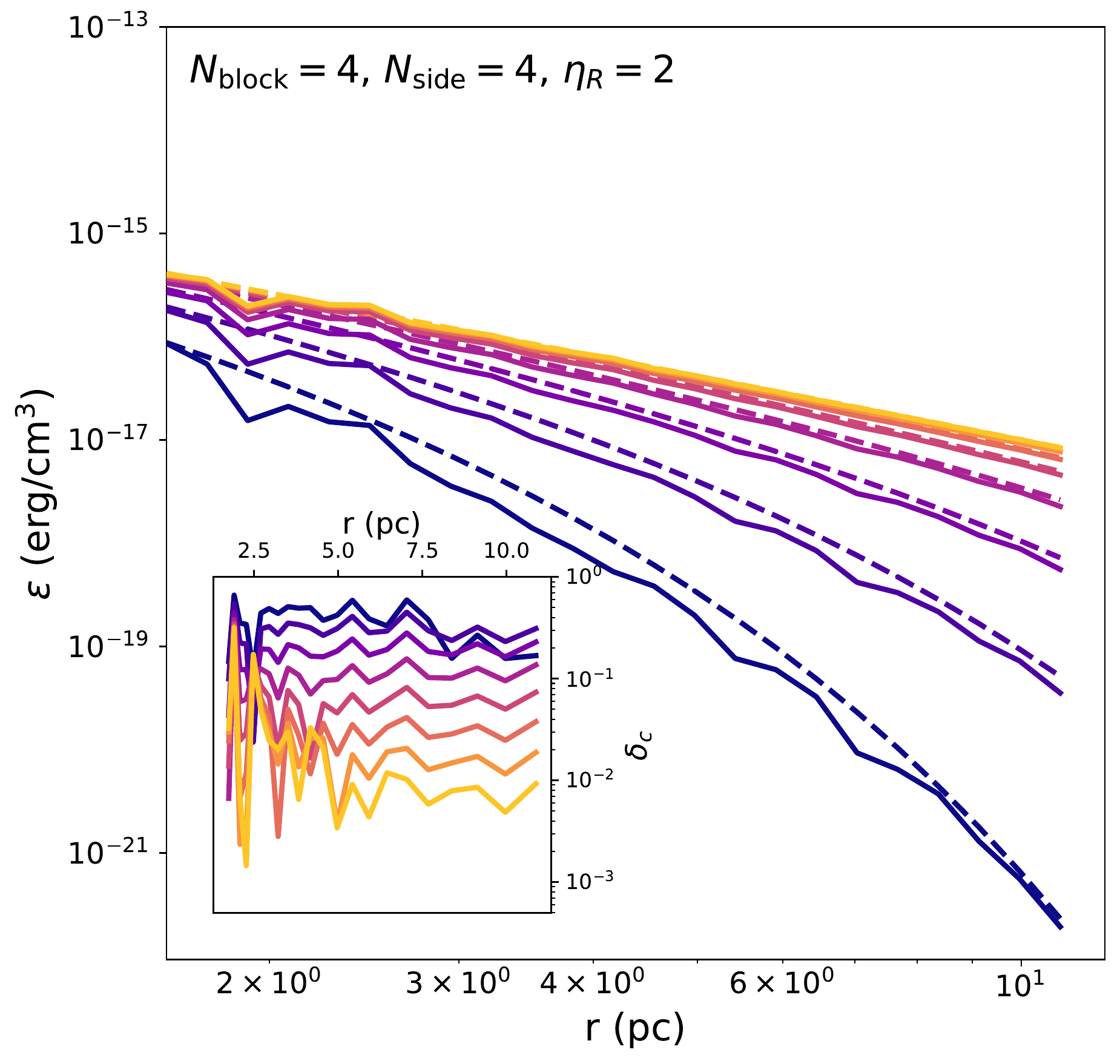} &
        \includegraphics[width=0.45\textwidth]{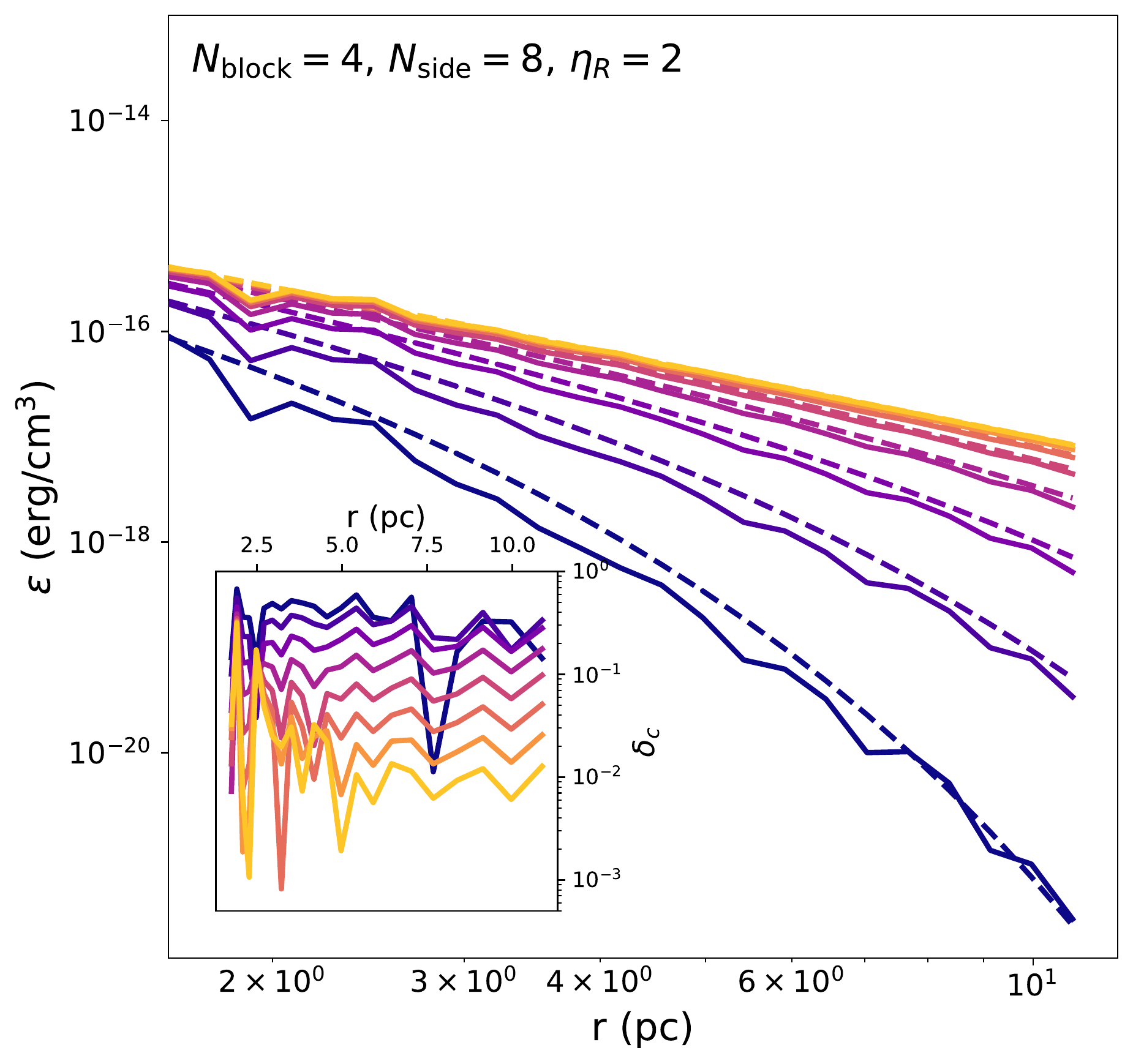} \\
        \includegraphics[width=0.45\textwidth]{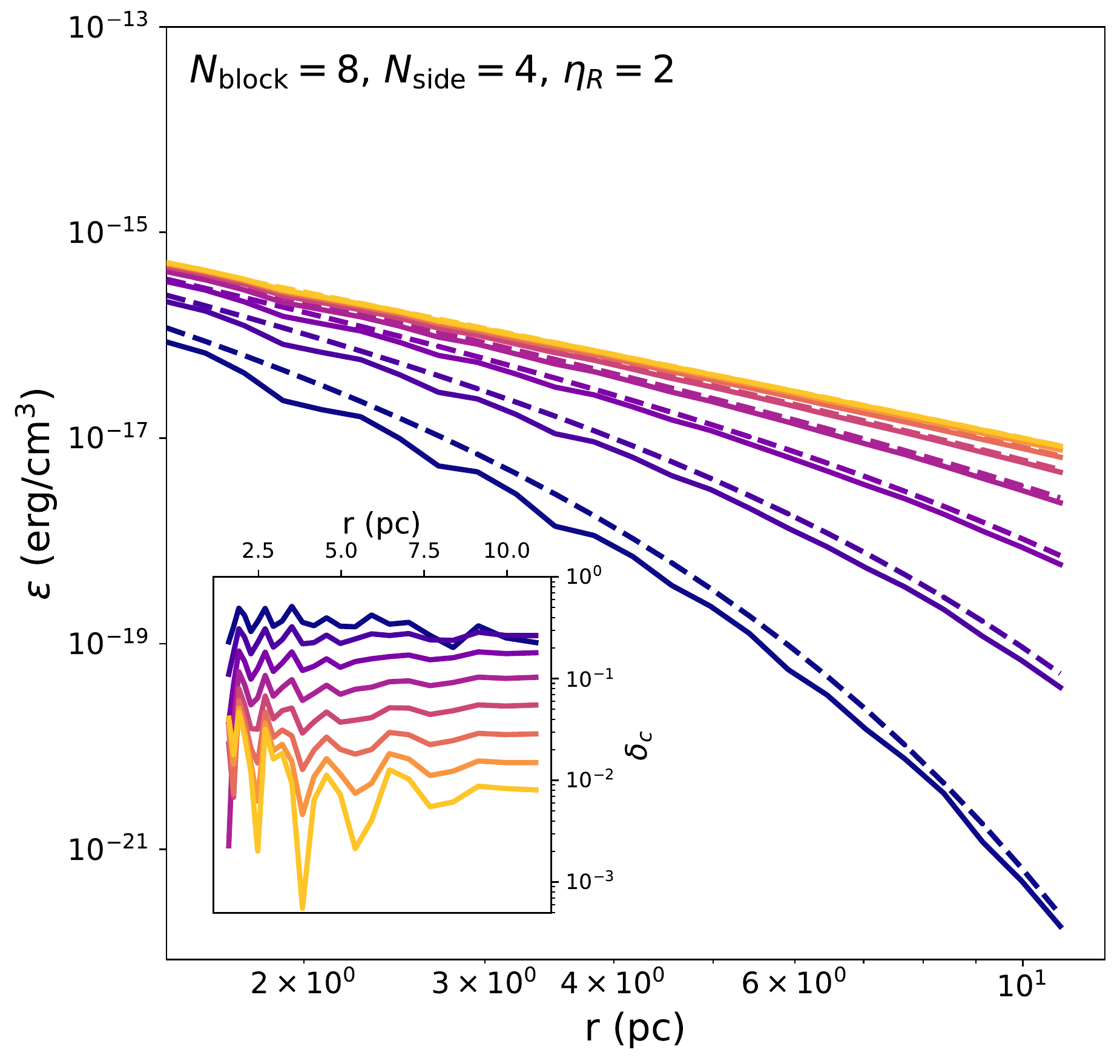} & 
        \includegraphics[width=0.45\textwidth]{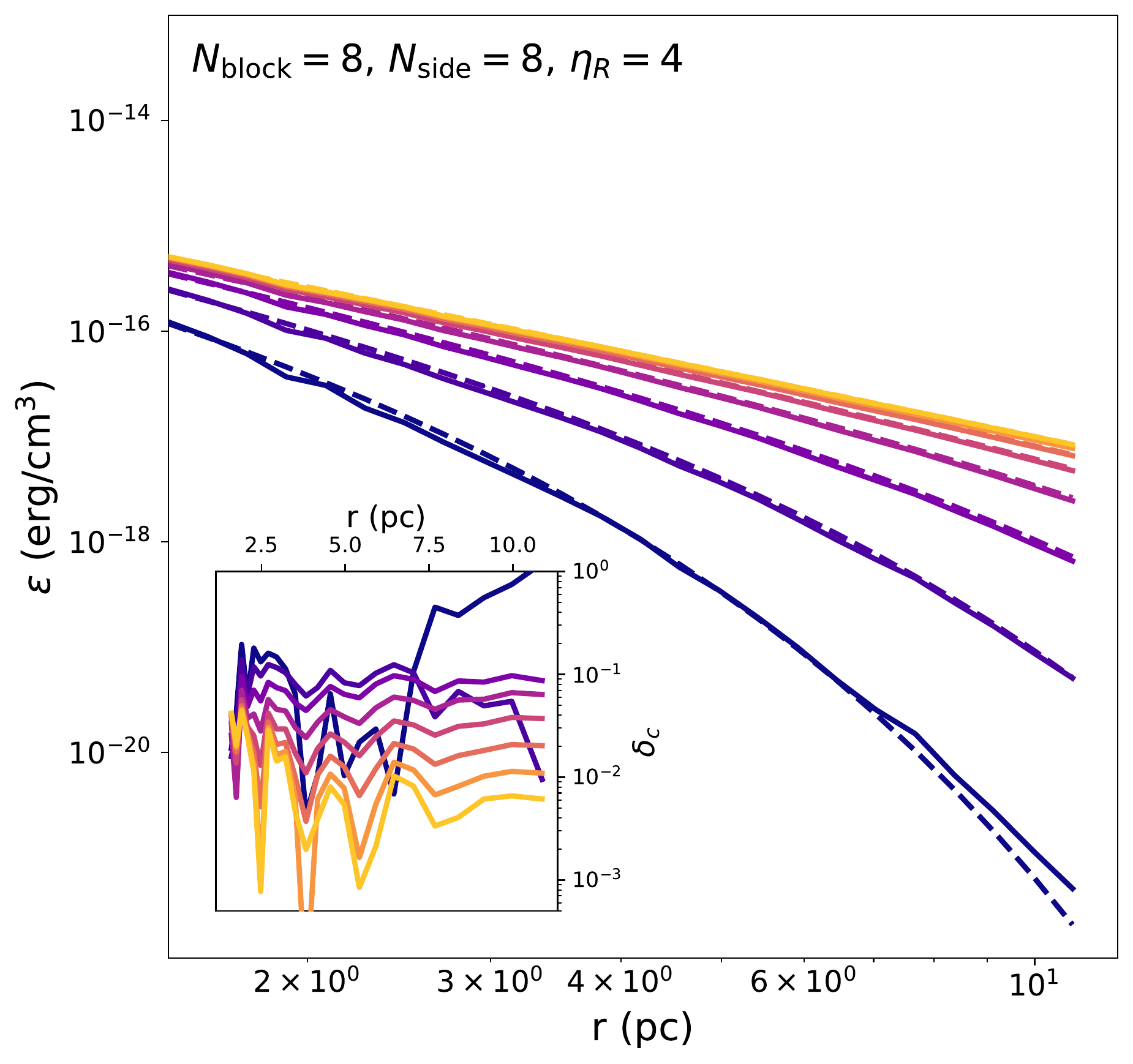}
    \end{tabular}
    \caption{\label{fig:radProfs}. Energy density versus radius for the different model parameters, annotated in the top left of each subfigure. Inset: Relative error, $\delta_c$, of the numerical solution against the analytic solution as a function of radius from the source.}
\end{figure*}

Figure \ref{fig:radProfs} shows the performance of {\sc XRayTheSpot} for a range of parameters, exploring both low- and high- spatial and ray resolutions. We find that grid resolution is the primary source of deviations at small radius, while the ray resolution increases the accuracy at larger radii. At large distances from the source, the solution tends to slightly under predict for low ray and angular resolution due to overestimation of the column density. At small radii, the solution over-predicts the resulting flux. For optically thin radiation bins, the solution almost exactly matches the analytic. Therefore, the deviations come about due to mapping the mass from the cells and tree nodes onto the rays using the kernel.

For science uses, the number of blocks, $N_{\rm block}$, is determined by the necessary resolution to resolve crucial gas dynamics (e.g. the Jeans length for gravity simulations). Increasing both $N_{\rm side}$ and $\eta_R$, while producing more accurate radiation transfer solutions, leads to substantially higher computational costs. Table \ref{tab:radTime} shows the computational time per processor for the initialization and per ray trace step for the models in Figure \ref{fig:radProfs}. Between the lowest and highest accuracy tests, $(N_{\rm block}, N_{\rm side}, \eta_R) = (4, 2, 2)$ and $(8,8,4)$, respectively, the increase in cost of the initialization and ray trace was a factor of $\approx$ 40 and $\approx$ 130, respectively. The time for the raytrace is dominated ($\geq 95\%$) by the tree walk. We find using $N_{\rm side} = 4$ is the best balance of time and accuracy.

\begin{table}
    \centering
    \begin{tabular}{ccccc}
        $N_{\rm block}$ & $N_{\rm side}$ & $\eta_R$ & Initialization (s/proc) & Evolution (s/proc) \\
        \hline
        4 & 2 & 2 & 2.3 & 1.3 \\
        4 & 4 & 2 & 5.5 & 3.6 \\
        4 & 4 & 4 & 6.1 & 4.6 \\
        4 & 8 & 2 & 22.1 & 14.0 \\
        8 & 4 & 2 & 17.4 & 29.3 \\
        8 & 8 & 4 & 94.0 & 167.2  
    \end{tabular}
    \caption{\label{tab:radTime}Timing for the pont source test for the different runs in Figure \ref{fig:radProfs}. Each row gives the model parameters of $N_{\rm block}$, $N_{\rm side}$ and $\eta_R$ and the time in seconds per processor for the initialization of the tree and a ray trace step.}
\end{table}

\subsection{Shadow Test}
\begin{figure}
    \centering
    \includegraphics[width=0.5\textwidth]{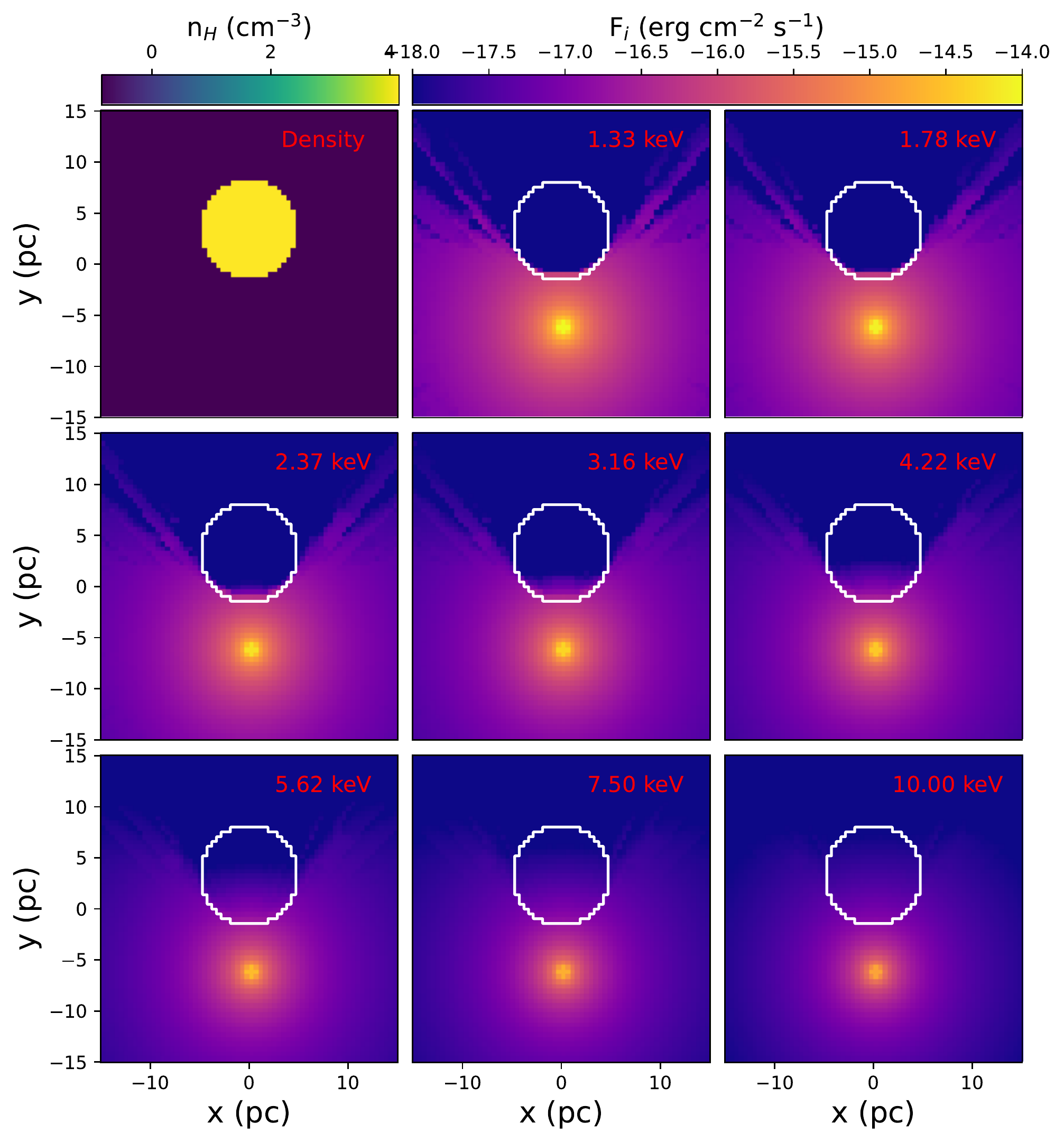}
    \caption{\label{fig:shadow}Shadow test, consisting of a point source illuminating a constant density core. Top left corner: Number density distribution for a z-axis slice. Others: X-ray flux in the given energy band for a z-axis slice using $N_{\rm block} = 8$, $N_{\rm side} = 4$, $\eta_R = 2$.}
\end{figure}

\begin{figure}
    \centering
    \includegraphics[width=0.5\textwidth]{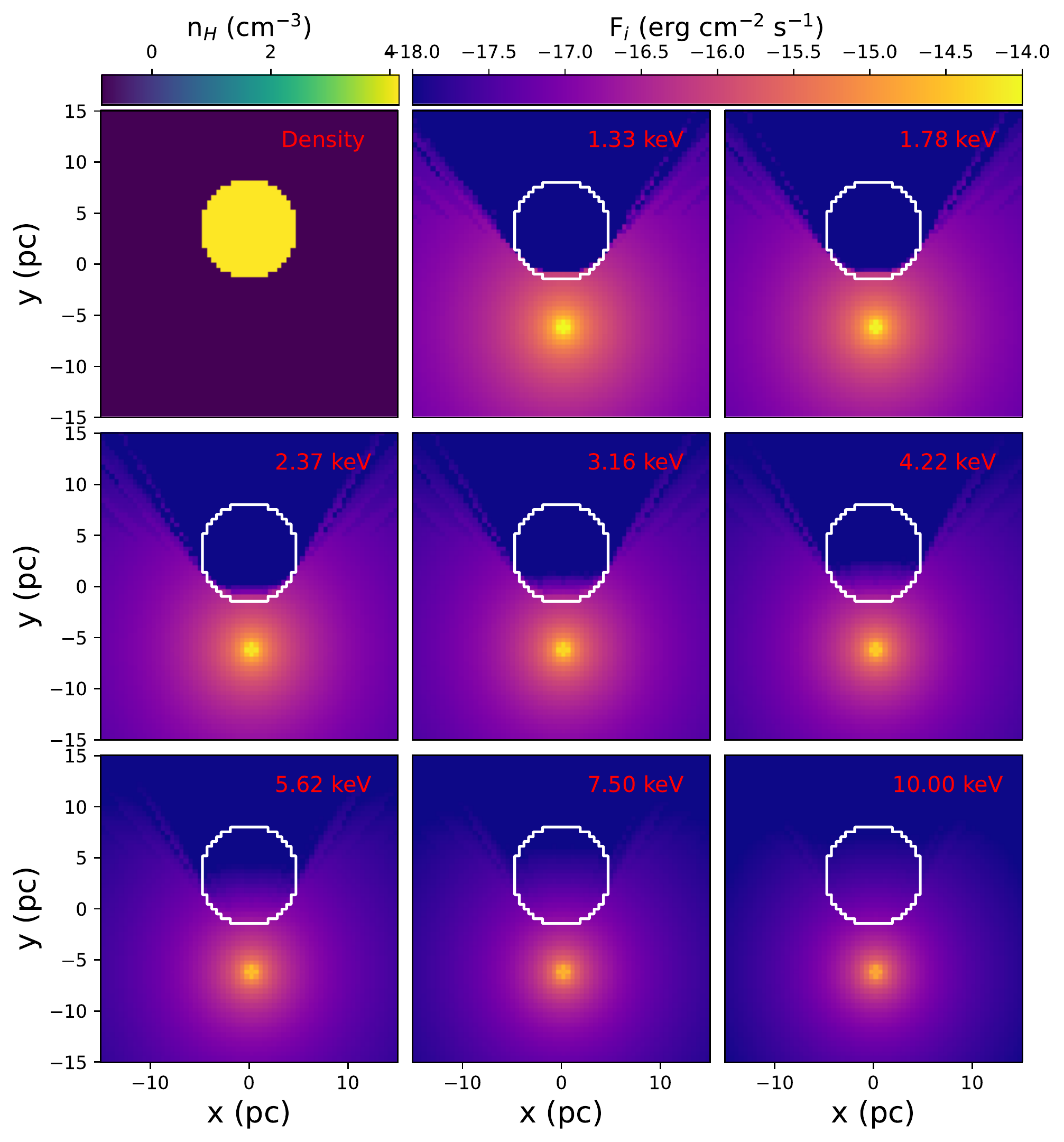}
    \caption{\label{fig:shadowHiRes}Same as Figure \ref{fig:shadow}, but with $N_{\rm block} = 8$, $N_{\rm side} = 8$ and $\eta_R = 4$. }
\end{figure}

Our next test is a shadow test to verify the solution of the solver when sources are placed near dense regions. Here, we have a $L_X = 10$ L$_{\odot}$ source with a spectrum, $dL/dE \propto E^{-2}$, placed near a dense core with a hydrogen-nuclei number density of $n_{\rm H} = 10^3$ cm$^{-3}$. We consider radiation between 1 - 10 keV, moving from optically thick bands to optically thin. Figures \ref{fig:shadow} and \ref{fig:shadowHiRes} show the results of this test, for both low- and high- ray resolution which use ($N_{\rm block} = 8$, $N_{\rm side} = 4$, $\eta_R = 2$) and ($N_{\rm block} = 8$, $N_{\rm side} = 8$, $\eta_R = 4$), respectively. For both cases, the test reveals the expected results that the low-energy X-rays are absorbed by the dense core and this creates a wide-angle shadow, while higher energy X-rays are barely attenuated, producing smaller to no shadows. The high-ray resolution test also shows the expected drop in ray-tracing artifacts.

\begin{figure}
    \centering
    \includegraphics[width=0.45\textwidth]{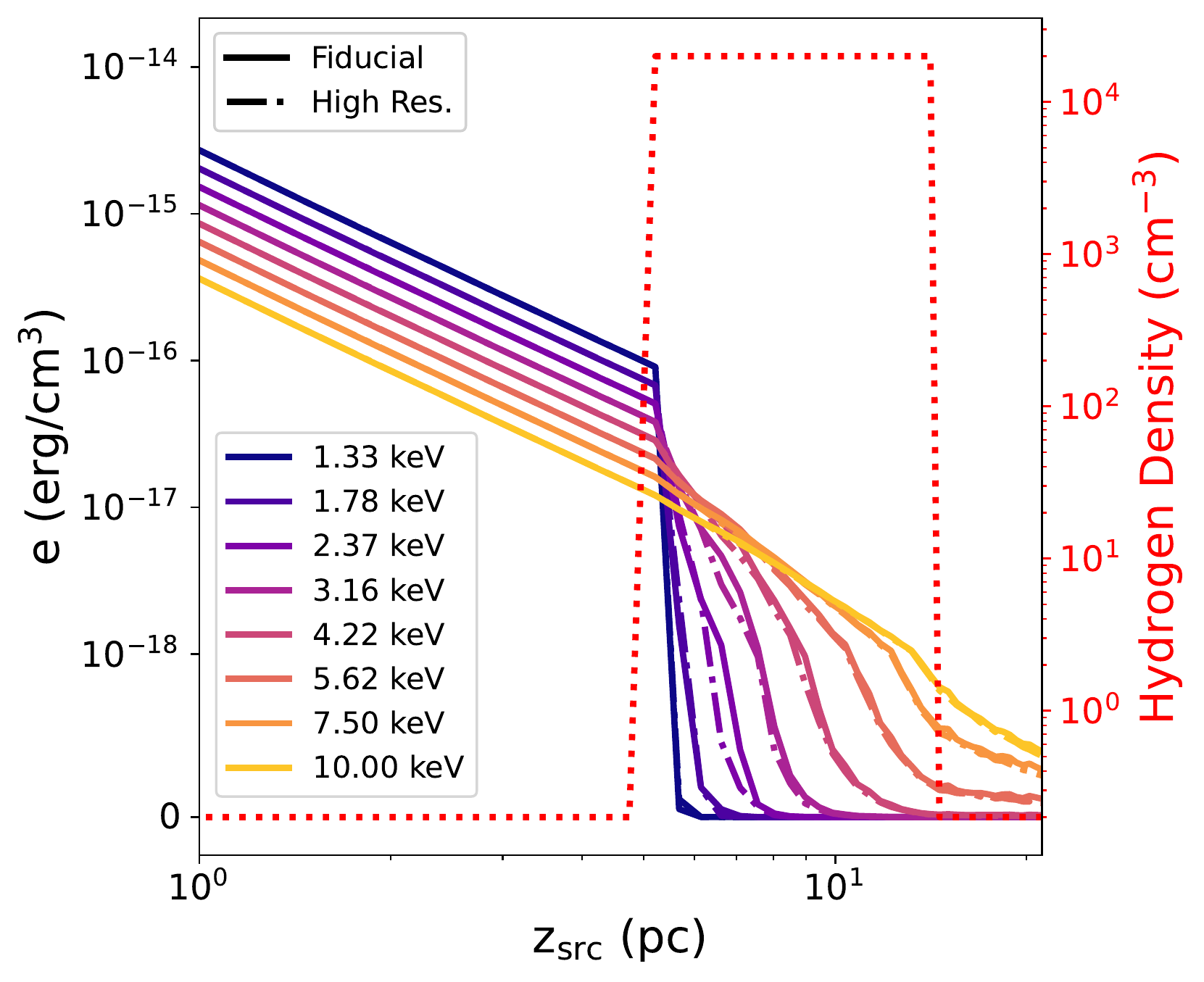} 
    \caption{\label{fig:shadowprof}X-ray energy density versus distance along the z-axis from the source for the shadow test. The solid line uses the ray resolution in Figure \ref{fig:shadow} and the dashed-dot uses the ray resolution in Figure \ref{fig:shadowHiRes}. The dotted red line shows the hydrogen nuclei density highlighting the location of the high-density blob.}
\end{figure}

Figure \ref{fig:shadowprof} shows a one-dimensional cut along the z-axis from the source through the dense blob of gas for the two ray resolutions compared. The figure shows that higher ray resolution leads to a smoother attenuation of the flux for the optically thick, lower energy bins while there is very little change for higher energy bins which are substantially less attenuated. This is most pronounced for the $E_c = 1.33$ keV bin

\subsection{Benchmark against Cloudy}
The final benchmark tests the X-ray radiation coupling to the thermochemistry. We place a point source with a physical size of $10^{16}$ cm and total X-ray luminosity, $L_{\rm XR} = 10^{36}$ erg s$^{-1}$, and a luminosity spectrum $dL/dE \propto E^{-2}$ between 1 - 10 keV in a (1.3 pc)$^3$ volume filled with a gas number density of $n_{\rm H} = 10^3$ cm$^{-3}$, $N_{\rm side} = 4$, and $\eta_R = 2$ and compare with a one-dimensional model using the {\sc Cloudy} code. {\b We also use a cosmic-ray ionizaiton rate of $3\times10^{-17}$ s$^{-1}$ and a gas and dust temperature floor of 3 K to mimic the inclusion of a cosmic microwave background.} The volume and resolution were chosen such the inner XDR is resolved by $\approx 20$ cells while the optically thick regime is also traced. In particular, we use 7 maximum adaptive-mesh resolution levels refining on the density and temperature, such that the maximal resolution is $1.3\times10^{-3}$ pc. The one-dimensional {\sc Cloudy} model used a ``sphere'' geometry, an input power-law spectrum between 1 - 10 keV for the X-ray radiation source, a cosmic microwave background and a cosmic-ray ionization rate of $3\times10^{-17}$ s$^{-1}$. Further, we turn off grain physics, induced radiative processes, radiation pressure, radiation scattering, outward line radiation transfer and molecule freeze-out, since the {\sc Flash} simulations do not have these processes. Finally, we set refractory metal abundances to zero, with the exception of silicon which {\sc Flash} uses as the proxy for metals for the chemistry (as described above). The {\sc Cloudy} script used is shown in Appendix \ref{sec:cloudApp}.

Since we using uniform-spaced grids, even with substantial AMR levels, it is in practice difficult to fully capture sharp thermochemical transition regions, such as that shown below as captured by {\sc Cloudy}. Further, the source encompasses several cells at the highest resolution, rather than an infinitely small point source. Capturing such ionization and dissociation fronts entirely is numerically intensive and generally requires the use of one-dimensional models tailored to do so (as with {\sc Cloudy}). 

\begin{figure*}
    \centering
    \includegraphics[width=0.95\textwidth]{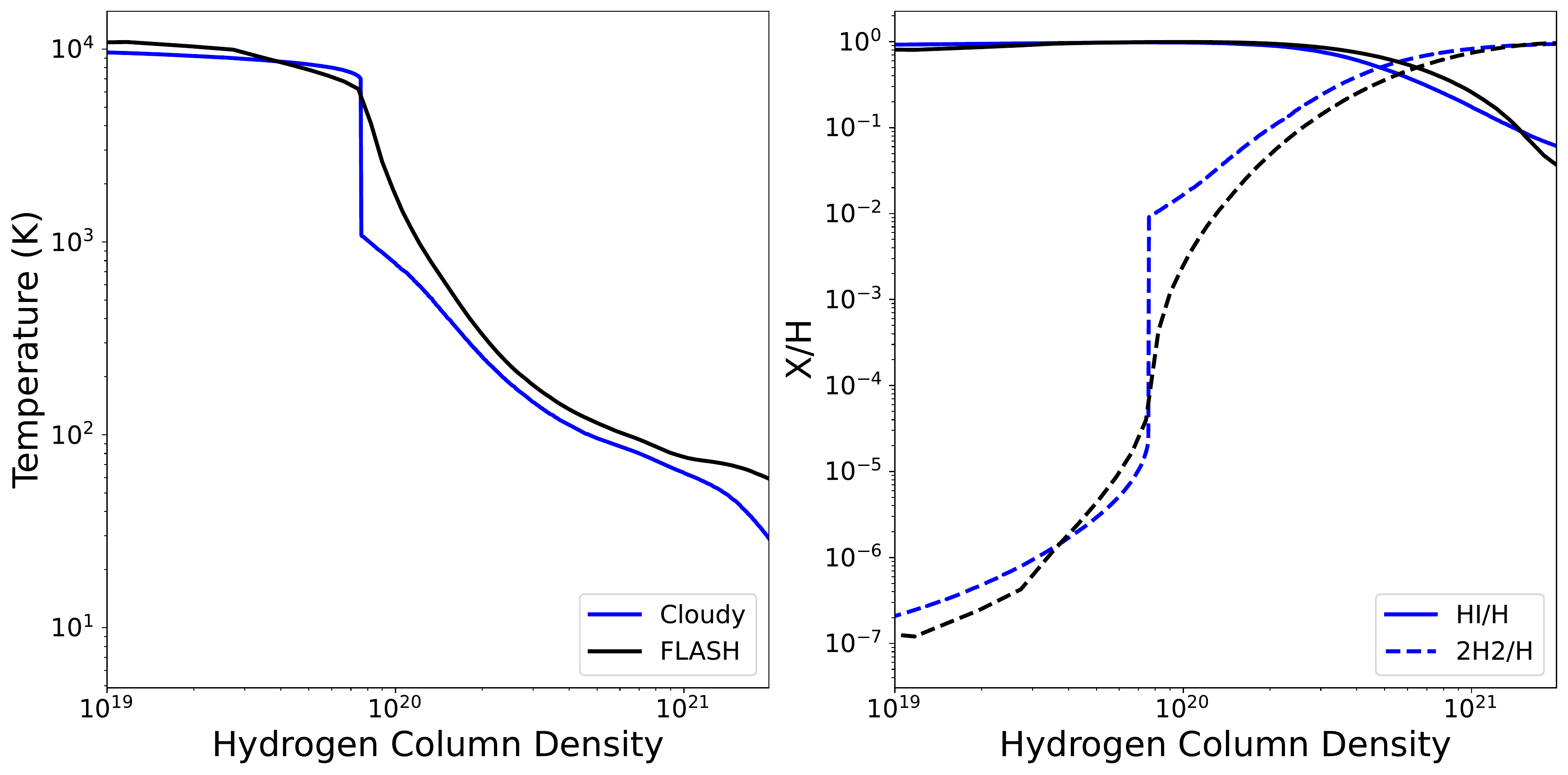}
    \caption{\label{fig:cloudyComp}{\sc Flash} vs {\sc Cloudy} benchmark. Left: Temperature versus hydrogen column density from the central point source for {\sc Flash} (black) and {\sc Cloudy} (blue). Right: Atomic (solid) and molecular (dashed) hydrogen abundances versus total hydrogen column density from the source, where H$_{\rm tot}$ = H$^+$ + H + $2$H$_2$. }
\end{figure*}

Figure \ref{fig:cloudyComp} shows the result of this benchmark. Near the source, the temperature and chemistry solutions well match the {\sc Cloudy} solution. The {\sc Flash} and {\sc Cloudy} solutions qualitatively reproduce the chemical structure, although due to the larger cell-size of the {\sc Flash} grids, the sharp HI transition seen at $N(H) \approx 5\times10^{20}$ cm$^{-2}$ is not fully captured and is instead smoothed over a few cells. The temperature solutions agree within a factor of a few. However, {\sc Cloudy} solves the line cooling and level excitations in a much more robust manner than the included {\sc Flash} thermochemistry, including a full non-equilibrium solution with many more electronic and ionization states. Such inclusions though are not numerically feasible for in-situ thermochemistry in three-dimensional MHD simulations. Further, {\sc Cloudy} solves the full radiation transfer solution from radio through X-ray radiation with substantially more bins. 

\begin{figure}
    \centering
    \includegraphics[width=0.5\textwidth]{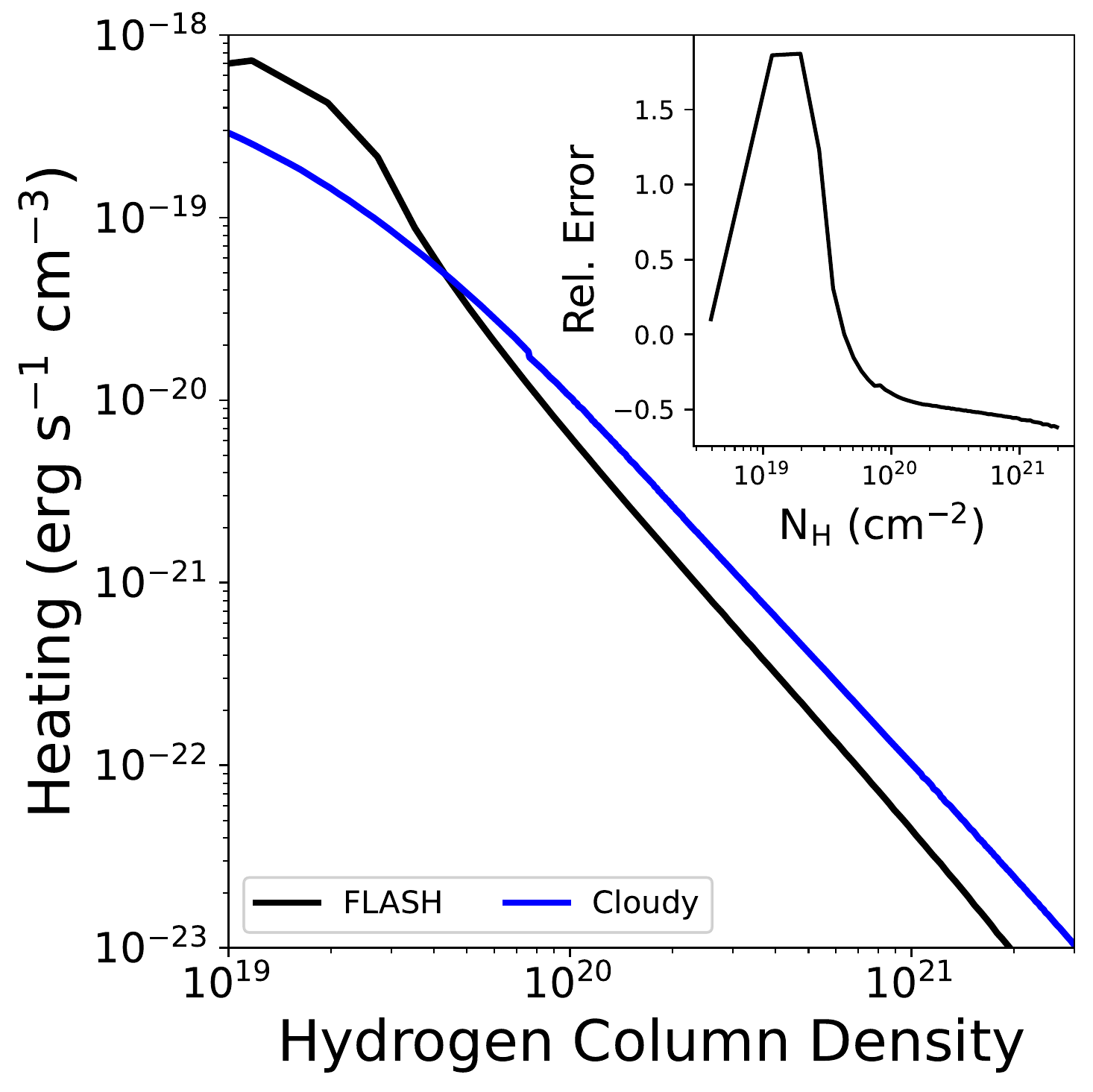}
    \caption{\label{fig:cloudyHeat} {\sc Flash} vs {\sc Cloudy} heating benchmark. Total heating from {\sc Cloudy} (blue) versus X-ray heating from {\sc Flash}. Inset: Relative difference between these. }
\end{figure}

Figure \ref{fig:cloudyHeat} shows a comparison of the heating rates. For {\sc Flash}, the X-ray heating is included via
\begin{equation}
    \Gamma_X = \eta_X n_{\rm H} H_{\rm X},
\end{equation}
where $\eta_X$ is the heating efficiency \citep{dalgarno1999}. We calculate this heating term in post-processing for the {\sc Flash} runs, as the chemistry solver does not store this in real time in the simulation. For {\sc Cloudy}, the heating term output does not delineate heating caused by X-ray radiation. However, since we include only a low cosmic-ray background with no other primary radiation heating source, the total heating rate will provide a close approximation. We find that our heating term is, on average, within a factor of two of the total heating rate computed by {\sc Cloudy}.

\begin{figure}
    \centering
    \includegraphics[width=0.5\textwidth]{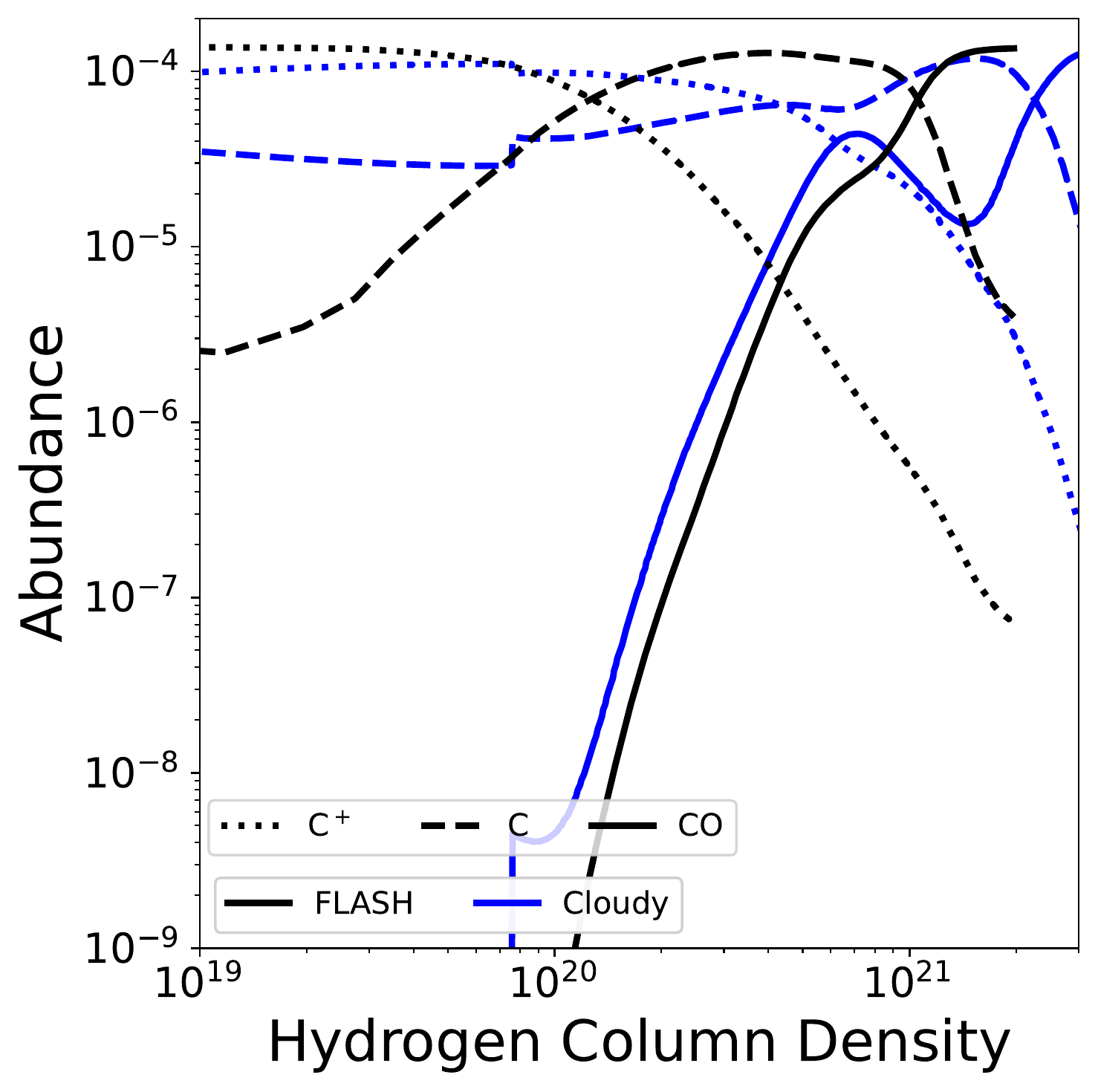}
    \caption{\label{fig:cloudyCarb} {\sc Flash} vs {\sc Cloudy} benchmark. Abundances of \ce{C+} (dotted), \ce{C} (dashed) and \ce{CO} (solid) versus total hydrogen column density from the central point source for {\sc Flash} (black) and {\sc Cloudy} (blue).}
\end{figure}

Figure \ref{fig:cloudyCarb} compares the relative abundances in the so-called ``carbon cycle'', \ce{C+ -> C -> CO}, between the two codes. The results broadly agree, with {\sc Flash} showing an enhanced neutral Carbon and rapid production of CO, which is a result of the greatly reduced network \citep{Glover2010, gong2017}. {\sc Cloudy} uses a significantly larger network with detailed line radiation transfer for self-shielding (important for \ce{H2} and \ce{CO}) while we use the average column densities of {\sc H2} and {\sc CO} to compute the self-shielding factor \citep{Walsh2015, wunsch2018}. Given the constraints of these physics, the found solution is deemed to be adequate and matches the overall trends as determined by {\sc Cloudy}. 

\section{Protostellar Disk}\label{sec:disk}
Evolved protostellar objects, in particular Class II objects in which the lack of a surrounding gaseous envelope leaves the central protostar and disk exposed, are known to be X-ray emitters. These X-rays can become important for disk dynamics and planet formation \citep[e.g.][]{Ercolano2008b, Mohanty2013}. For these stars, the X-ray emission is thought to come from a combination of accretion and magnetospheric emission \citep{Hartmann2016}. As a first test science case, we model the X-ray radiation transport from a central protostar into a protostellar disk. 

The surface density follows from the often used truncated power-law \citep[e.g.][]{lyndenbell1974, andrews2011, cleeves2016}:
\begin{equation}
    \Sigma_g(R) = \Sigma_c \left ( \frac{R}{R_c}\right )^{-\alpha} \exp \left [ -\left ( \frac{R}{R_c}\right )^{2-\alpha}\right]
\end{equation}
between an inner and outer radius, $R_{\rm in}$ and $R_{\rm out}$, respectively, $R_c$ is the critical radius where the surface density distribution becomes exponential, $\alpha$ is the power law index and $\Sigma_c$ is the characteristic surface where the disk transitions to an exponential profile. For the initial conditions, we assume the gas is in hyrostatic equilibrium, such that the density follows
\begin{equation}
    \rho_g(R, z) = \frac{\Sigma_g(R)}{\sqrt{2\pi} h} \exp \left [ -\left ( \frac{z^2}{2h^2}\right )\right ]
\end{equation}
where $h = c_s/\Omega$ is the disk scale height, $c_s = \sqrt{\frac{\gamma k_b T_g}{\mu m_{\rm H}}}$, $k_B$ is Boltzmann's constant, $\gamma = 5/3$ is the adiabatic index, $T_g$ is the gas temperature, $\mu = 2.33$ is the mean mass per particle for molecular gas, $m_H$ is the mass of the hydrogen atom, $\Omega = \frac{3}{4}\sqrt{\frac{GM_*}{R^3}}$ is the Keplerian rotational frequency and $M_*$ is the mass of the central protostellar object. For this fiducial test, we set $M_* = 0.7$ M$_{\odot}$, $\Sigma_c = 64$ g cm$^{-2}$, $R_c = 100$ AU, $\alpha = 1$. The temperature profile is given by
\begin{equation}
    T(R) = {\rm max}\left [T_0 \left ( \frac{R}{1 AU} \right )^{-0.5}, 10 {\,\, \rm K}\right],
\end{equation} 
where we fiducially take $T_0 = 50$ K. The disk is initialized to be rotating in Keplerian motion around the central protostellar object. We assume the disk is magnetized with an initial toroidal field such that the ratio of the magnetic to thermal pressure, $\mu_M = 10^{-5}$. We simulate the domain in a 240 AU box with a maximal resolution of 1 AU.

The central protostar is put in by hand, with active accretion. For the X-ray emission, we assume an accretion floor of 10$^{-9}$ M$_{\odot}$ yr$^{-1}$, similar to rates observed in young stellar objects \citep[e.g.][]{Ingleby2013}. The simulation is run using the Bouchut-5 MHD solver, gravity, and {\sc XRayTheSpot}. The X-ray emission is derived by assuming there is an accretion shock, with properties following ``hot spot'' accretion \citep{Hartmann2016} with accretion columns filling 10\% of the protostar surface, thermally emitting X-ray emission. The thermal X-ray emission is computed using a one-temperature Raymond-Smith plasma model \citep{raymond1977}. The implementation of a coronal model is left for a future work.

\begin{figure*}
    \centering
    \includegraphics[width=0.95\textwidth]{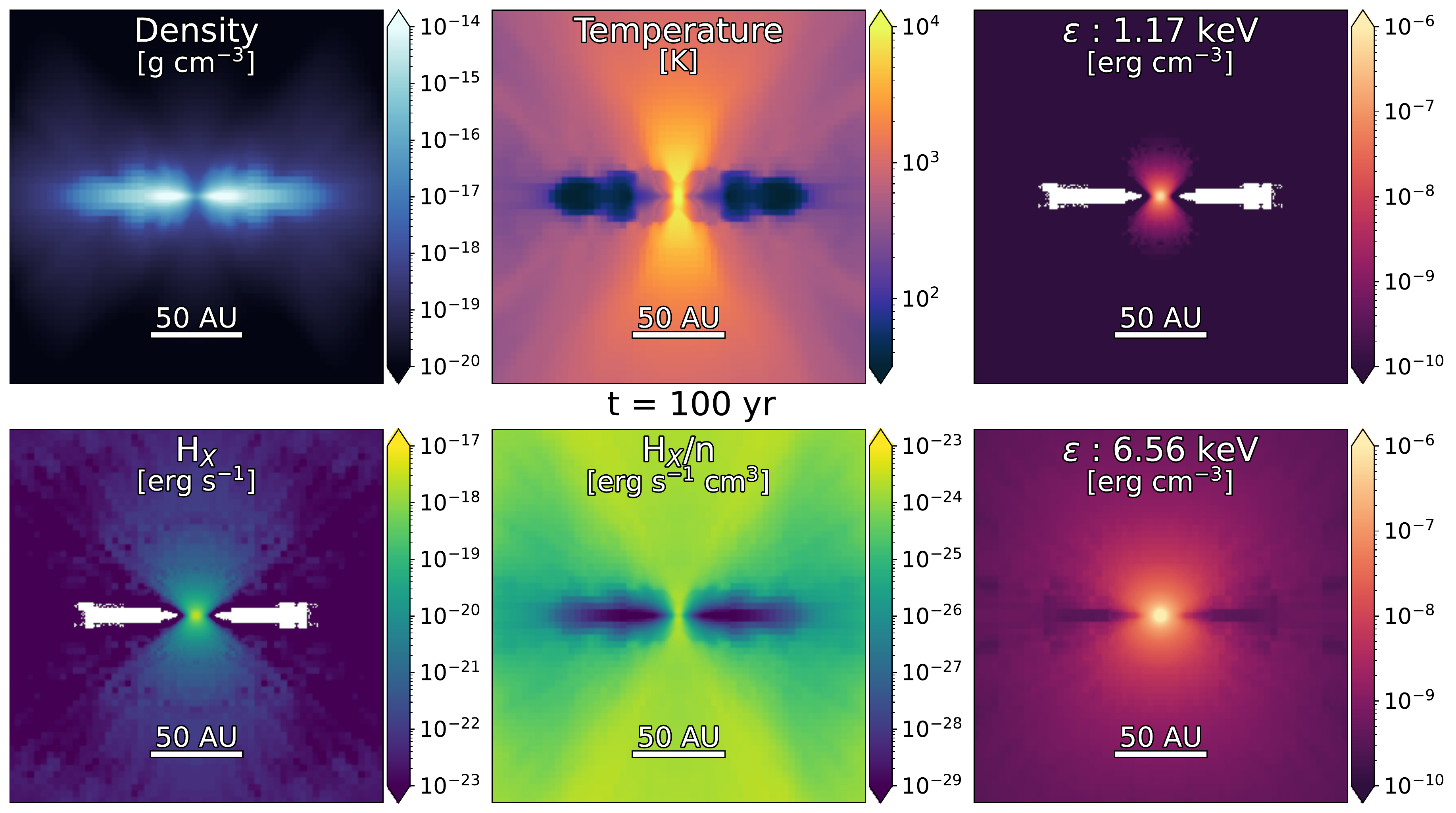}
    \caption{\label{fig:disk} Protostellar disk example case usage. Top row: Slice plots at $z = 0$ for the density (left), gas temperature (middle) and 1.17 keV radiation energy density. Bottom row: X-ray heating rate, H$_x$ (left), H$_x$/n diagnostic term (middle) and 6.56 keV radiation energy density.}
\end{figure*}

Figure \ref{fig:disk} shows a slice of the density, gas temperature, X-ray emission at 1.17 keV (1st bin) and 6.56 keV (8th bin), the heating rate per H nucleus, $H_x$, and $H_x/n$, which is often used as a diagnostic for the importance of the X-ray heating \citep{wolfire2022}. We find that the lowest energy X-rays are all absorbed near the protostar or escape through the outflow. However, the harder X-rays at 6.56 keV are able to permeate much of the domain. The $H_X$ and $H_x/n$ slices clearly show that the disk midplane is left relatively unheated by the X-rays, although the X-rays become important in the cavity and outer disk regions. In particular, most of the cavity exhibits very warm gas, even with only X-ray emission included, due to the rapid absorption of soft X-ray emission. The cavity heats to temperatures exceeding $10^4$ Kelvin, potentially becoming bright in hydrogen recombination lines. The inclusion of EUV radiation will heat the diffuse gas further, along with further ionizing the surrounding low-density cavity.

\section{Molecular Cloud}\label{sec:mcsim}
We present an example application for {\sc XRayTheSpot}, to demonstrate how all the different {\sc TreeRay} energy bands work together: a virialized, magnetized turbulent cloud. We consider a 2 pc region of a molecular cloud resolved with 256$^3$ cells. We produce an initial turbulent field by stirring the domain with a flat power spectrum between the largest wave modes $k = 1...3$ for 10 crossing times at a velocity dispersion of 0.72 km s$^{-1}$, consistent with the observed linewidth-size relationship \citep{McKee2007}. During the stirring, we use periodic boundary conditions and chemistry to achieve more accurate initial conditions for the abundances before collapse. The choice of stirring for 10 crossing times is to ensure the chemistry has reached a more quiescent state, with the kinetic energy spectrum generally being reached after two crossing times \citep{Federrath2010}. We assume the cloud is nearly virialized, such that the virial parameter
\begin{equation}
    \alpha \equiv \frac{5 \sigma^2 R}{G\rho L^3} = 2
\end{equation}
where $R = L$ is the box length, resulting in $\rho = 5\times10^{-21}$ (g cm$^{-3}$) and a total box mass of $M = 590$ M$_{\odot}$. Before stirring, we initialize a magnetic field in the $z$-axis with a magnitude such that the plasma beta,
\begin{equation}
    \beta \equiv \frac{\rho c_s^2}{B^2/8\pi} = 10^3.
\end{equation}
After the turbulence is initialized, gravity and source particles (stars) are turned on and the boundary conditions are changed to ``diode'' such that gas can flow out of the domain. During the simulation, the cloud is irradiated by an FUV radiation field of $\chi = 1.7$ in units of the Habing field \citep{Habing1968}. The simulation is run using the chemistry described above, and all {\sc TreeRay} modules: 
\begin{itemize}
    \item {\sc OpticalDepth} for the external radiation field \citep{wunsch2018}. {\sc OpticalDepth} solves for the column density from a cell to the external boundary and attenuates a prescribes external radiation flux ($\chi = 1.7$). In this study, it is only used for the FUV radiation, while \citep{mackey2019} implemented the ability to include an impinging X-ray flux.
    \item {\sc OnTheSpot} for the EUV emission \citep{wunsch2021}. This module solves for UV-ionizing radiation from arbitrary sources and iterates to convergence. The UV photon flux is coupled to the thermochemistry to model photochemistry.
    \item {\sc RadPressure} to account for the thermal radiation and radiation pressure \citep{klepitko2022}. This module enables the inclusion of thermal radiation from point and diffuse sources and the resulting radiation pressure. The thermal radiation is included in the chemistry through radiative dust heating.
    \item {\sc XRayTheSpot}, described above.
\end{itemize}

Sink particles representing protostars are injected when the density exceeds $\rho_{\rm thresh} \ge 4.59\times10^{-18}$ g cm$^{-3}$. Further criteria are used: there are checks to ensure a local gravitational potential and a converging flow. The protostar evolution follows the \citet{offner2009} model and implemented in {\sc Flash} \citep{klepitko2022}. Protostellar emission consists of the intrinsic and accretion luminosities, where the total accretion luminosity is
\begin{equation}
    L_{\rm acc} = f_{\rm acc} \frac{GM_*\dot{M}_*}{R_*},
\end{equation}
where $M_*$ is the mass of the protostar, $\dot{M}_*$ is the accretion rate, $R_*$ is the protostar's radius and we take $f_{\rm acc} = 0.33$. The X-ray spectrum was computed by assuming hot-spot accretion, described above, which provides the temperature and the density of the accretion shocks near the protostellar surface \citep{calvet1998, Hartmann2016} and a single temperature plasma model \citep{raymond1977}. Due to the low resolution, we set a minimum of $\dot{M}_* = 10^{-9}$ M$_{\odot}$ yr$^{-1}$. This is needed since when the protostar particles first form, the burst of accretion blows out HII regions, and the low resolution inhibits resolving the proper structure around the cores. The infrared to EUV spectrum, used for {\sc RadPressure} and {\sc OnTheSpot} is computed assuming the emission is composed of two blackbodies: one for the intrinsic spectrum of the protostar at the photosphere, such that
\begin{equation}
T_* = \left ( \frac{L_*}{4\pi \sigma_{\rm sb}R_*^2}\right )^{1/4},
\end{equation}
which is provided by the protostellar evolution model, and another assuming the accretion luminosity is reprocessed primarily as a blackbody with temperature $T_{\rm acc}$, such that
\begin{equation}
T_{\rm acc} = \left ( \frac{L_{\rm acc}}{4\pi\sigma_{\rm sb}R_*^2} \right )^{1/4},
\end{equation}
where $\sigma_{\rm sb}$ is the Stefan-Boltzmann constant. Therefore, the total infrared luminosity from the protostar is described as
\begin{equation}
L_{*, \rm{IR}} = f_{*, {\rm IR}}(T_*) L_* + f_{\rm acc, IR}(T_{\rm acc}) L_{\rm acc}
\end{equation}
and the EUV luminosity as
\begin{equation}
L_{*, \rm{EUV}} = f_{*, {\rm EUV}}(T_*) L_* + f_{\rm acc, EUV}(T_{\rm acc}) L_{\rm acc}
\end{equation}
where $f_{\rm IR}(T)$ and $f_{\rm UV}(T)$ are the fraction of the blackbody emission in each of these bands ($E < 13.6$ eV and $13.6$ eV $\le E \le 100$ eV, respectively). The X-ray emission was computed assuming the ``hot-spot'' model, described above. While there may be some double counting of emission by treating the total spectrum in the two different methods, we find this impact is marginal as the X-ray emission generally accounts for only a small fraction ($\le 10$\%) of the total protostellar luminosity.

\begin{figure*}
    \centering
    \includegraphics[width=0.95\textwidth]{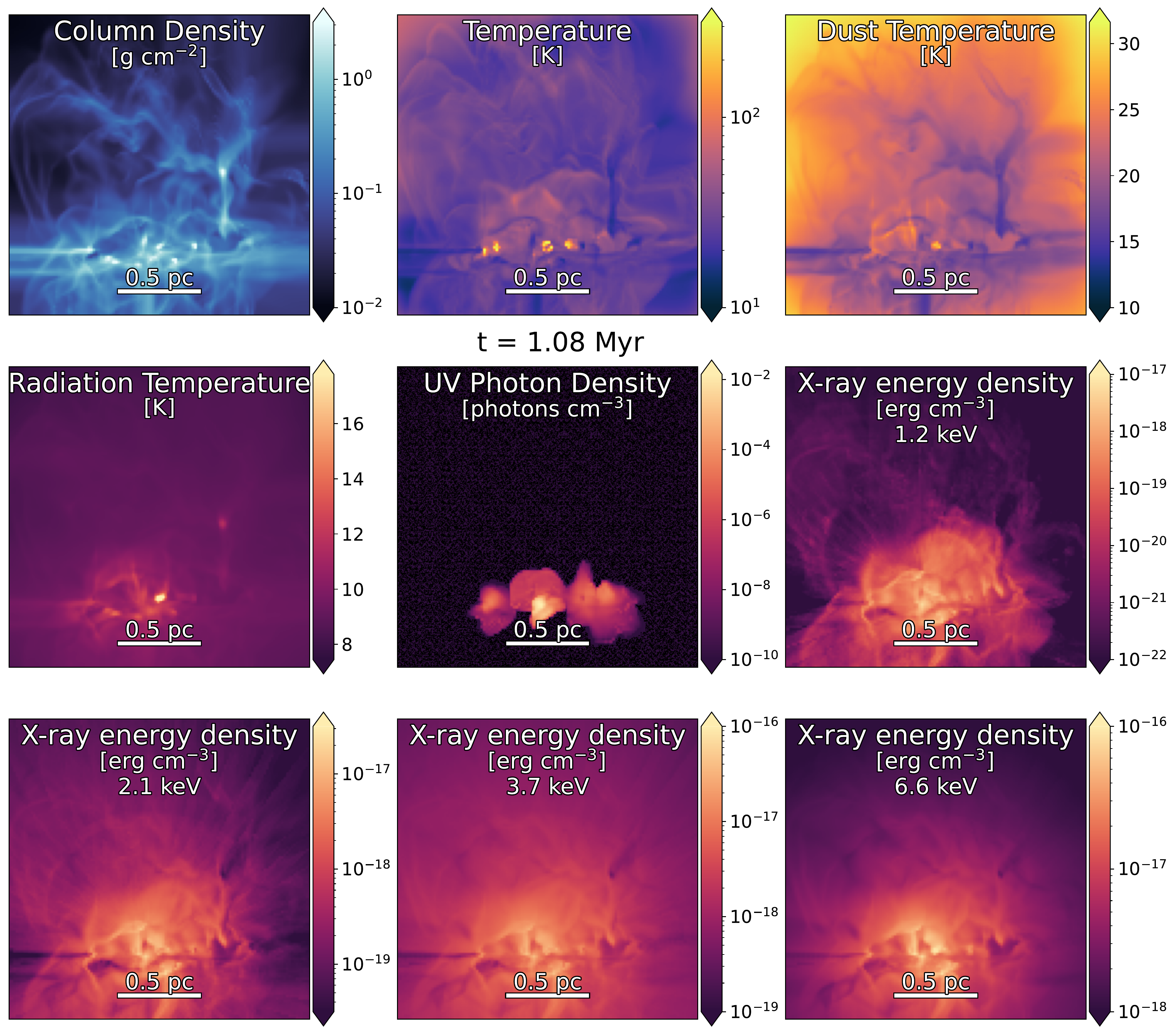}
    \caption{\label{fig:cloud} Panel plots highlighting the features of a 2 pc piece of a molecular cloud after $t = 1.08$ Myr of evolution. For all fields except the column density, the panel is showing the density-weighted projection. All projections are along the z-axis. The figure shows a simulated molecular cloud after 1 Myr of gravitational evolution including protostar sink particles and radiation feedback from infrared to X-rays. While the EUV radiation is rapidly absorbed (indicated by the black background color), the infrared and X-ray emission is able to penetrate much further into the cloud. }
\end{figure*}

Figure \ref{fig:cloud} shows the column density, and density-weighted projections of the gas and temperature, radiation temperature, EUV photon density and X-ray energy densities after $\approx$ 1 Myr of evolution with gravity. The star formation, as traced by  heated knots of gas, is occurring along a main filament structure. The high temperatures here are primarily caused by the EUV photons, which are rapidly absorbed in the nearby gas. The X-ray emission is found to be highly absorbed along the main filament structure, and instead traces out the more diffuse turbulent structure of the molecular cloud. As expected, higher energy X-ray bands showcase more extended emission with the brightest emission in the 3.7 keV band. In all X-ray bands, the turbulent structure of the molecular cloud is seen in the density-weighted integrated emission. 

\begin{figure*}
    \centering
    \includegraphics[width=0.95\textwidth]{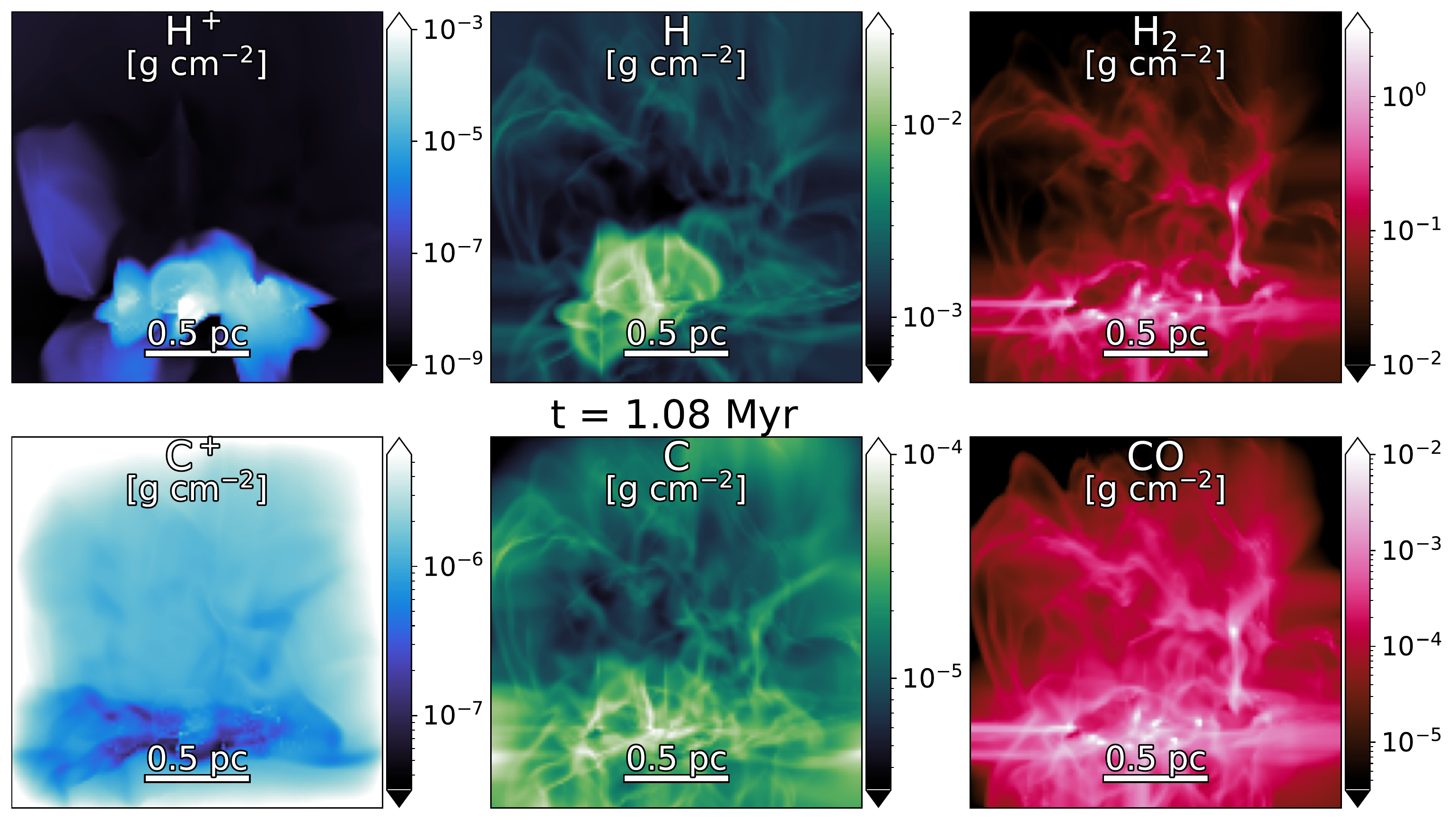}
    \caption{\label{fig:cloudchem} Panel plots highlighting the features of a 2 pc piece of a molecular cloud after $t = 1.08$ Myr of evolution. The panels show the integrated gas surface density of \ce{H+}, \ce{H} and \ce{H2} (top left, center and right) and \ce{C+}, \ce{C} and \ce{CO} (bottom left, center and right).}
\end{figure*}

\begin{figure}
    \centering
    \includegraphics[width=0.5\textwidth]{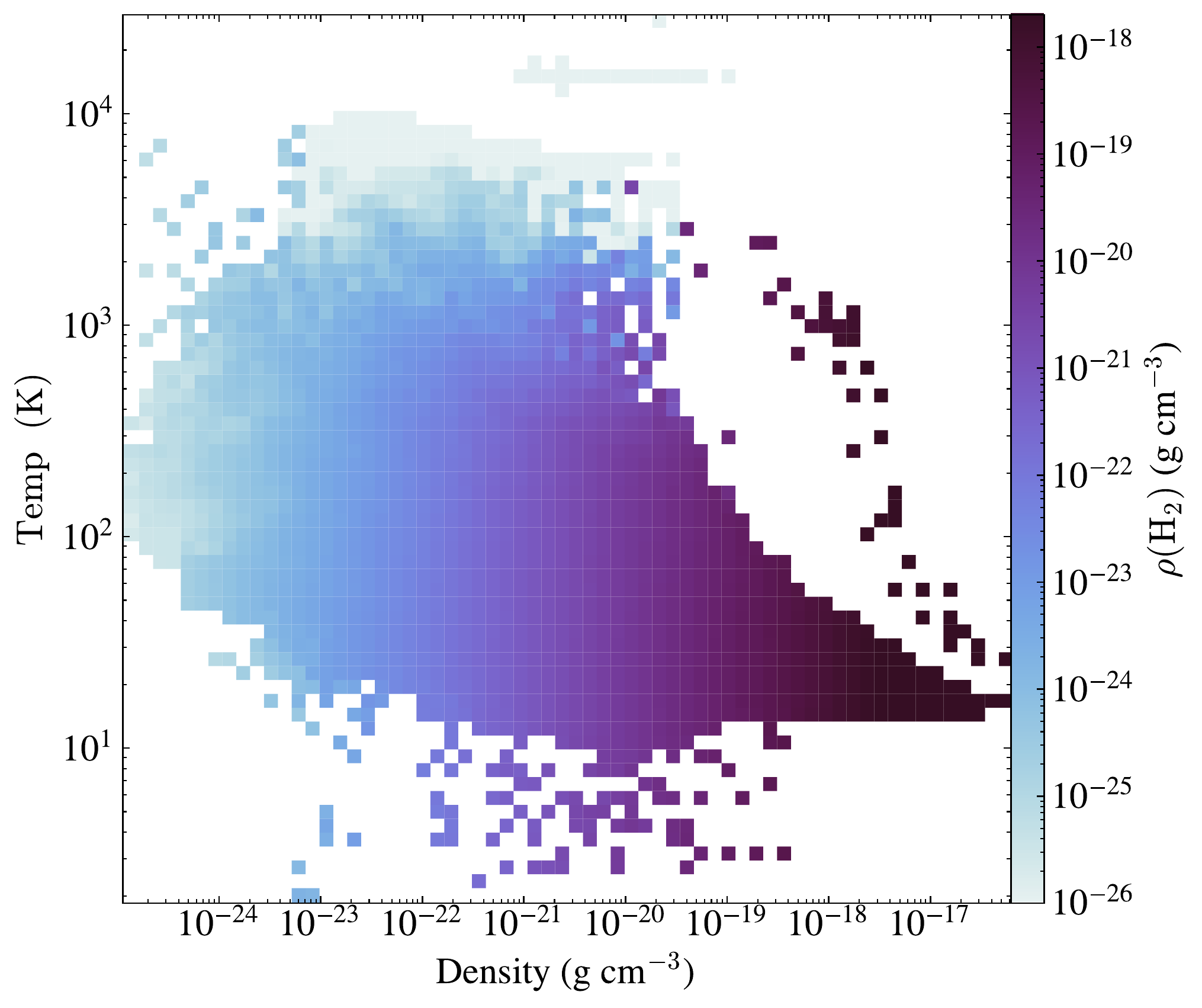}
    \caption{\label{fig:cloudphase} Density-temperature phase diagram, weighted by the \ce{H2} gas density. The phase diagram shows the typical qualitative behavior of self-gravitating turbulent molecular clouds, with a population of atomic gas at $T \approx 1.2\times10^4$ K and another population of dense, molecular gas with $10^2 \le T \le 10^3$, potentially resulting from X-ray heating.}
\end{figure}

Figure \ref{fig:cloudchem} shows the surface densities of atomic and molecular hydrogen, and atomic carbon and carbon monoxide species in the cloud and Figure \ref{fig:cloudphase} shows the \ce{H2} density-weighted density-temperature phase diagram. In this preliminary investigation, we see that there is an enhancement of neutral hydrogen roughly corresponding to where there are enhancements in the X-ray radiation field with some overlap of the UV-ionized \ce{H+}. Similarly, in the phase-diagram, there is a population of hot atomic gas, exceeded typical temperatures of HII regions \citep{Haid2018} and an additional population of dense, molecular gas at temperatures $10^2 - 10^3$ K. These regions are likely X-ray heated gas close to X-ray
emitting protostars, but a detailed investigation is beyond the scope of this work. This case study highlights the new capabilities of including protostellar radiative feedback from infrared to X-ray in star formation simulations.

\section{Discussion/Future Work}\label{sec:discuss}
We have presented the new X-ray radiation transfer module, {\sc XRayTheSpot} using the reverse ray-tracing scheme {\sc TreeRay} implemented in {\sc Flash} \citep{wunsch2021}. {\sc XRayTheSpot} enables an arbitrary number of point or diffusive sources of X-ray emission, and an arbitrary number and position of energy bins. The module uses temperature dependent cross sections assuming gas in thermal ionization equilibrium. However, the module is flexible enough such that the user can provide their own cross section data to be used. The module produces the expected behavior for X-ray point sources and shadow tests and is able to reasonably reproduce the thermochemistry compared to {\sc Cloudy}, despite the significantly simpler treatment of X-ray chemistry and grain-processes in {\sc Flash}. Hence the {\sc XRayTheSpot} module allows the inclusion of, for instance, time dependent feedback from protostars or X-ray binaries, or extended X-ray emission from hot, cooling gas with the X-ray radiation transport coupled to the hydrodynamics, chemistry and thermodynamics.

We demonstrated the utility of this module with two example science cases focusing on protostellar X-ray emission. First, we modelled the emission of an 0.7 M$_{\odot}$ protostar with an accretion rate of 10$^{-9}$ M$_{\odot}$ yr$^{-1}$ through a protostellar disk. We find that soft X-rays are rapidly absorbed at the disk surfance, with most of the emission escaping through the outflow cavity. However, harder X-rays are able to permeate the disk due to their significantly lower optical depth. The X-ray heating was also strong within the outflow cavity, with no X-ray heating towards the midplane of the disk, as expected. Second, we perform a low-resolution star formation simulation of a turbulent molecular cloud. In this simulation, protostars are self-consistently formed and the X-ray emission modelled on the fly. This simulation includes the entire range of different {\sc TreeRay} radiation modules: diffuse FUV ({\sc OpticalDepth} \citet{wunsch2018}), EUV ({\sc OnTheSpot} \citet{wunsch2021}), thermal radiation and radiation pressure ({\sc RadPressure} \citet{klepitko2022}) and X-ray emission from 1 keV to 10 keV. Since the X-ray emission in the simulation comes entirely from accretion onto the protostars, the X-ray emission is highly variable. Due to the lower resolution and the inclusion of ionizing radiation, the accretion occurs in bursts followed by the expansion of HII regions, which cut off accretion. With higher resolution, accretion may still be able to occur through disks, instabilities and more porous density structures. In future work, we will perform higher resolution simulations to model star formation including chemistry and radiation feedback across the electromagnetic spectrum.

In this work, we focus primarily on point sources. However, {\sc XRayTheSpot} makes no differentiation between point sources versus extended more diffusion emission. Future studies will include diffuse X-ray emission from cooling hot gas and shocked gas. Our module currently includes the computation of X-ray emission from accretion onto protostars, and future work will include X-ray models for more types of point sources such as X-ray binaries. The module presented in this work will allow the first-generation of simulations of star formation and galaxies with the inclusion of a wide range of X-ray sources.

\section*{Acknowledgements}
BALG and SWG acknowledges support by the ERC starting grant No. 679852 ‘RADFEEDBACK’. SWG and BALG thank the German Science Foundation (DFG) for funding through SFB956 project C5. We also thank the Regional Computing Center Cologne (RRZK) for hosting our HPC cluster, Odin, on which the simulations have been performed. RW acknowledges the support by project 20-19854S of the Czech Science Foundation and by the institutional project RVO:67985815. JM acknowledges support from a Royal Society-Science Foundation Ireland  \emph{University Research Fellowship} (20/RS-URF-R/3712) and an Irish  Research Council \emph{Starting Laureate Award} (IRCLA\textbackslash 2017\textbackslash 83). The authors thank Andre Klepitko for many helpful discussions. Andre Klepitko also implemented the protostellar evolution model into the code. The authors thank the anonymous referee for their comments which improved the clarity of this work. The software used in this work was in part developed by the DOE NNSA-ASC OASCR Flash Centre at the University of Chicago \citep{fryxell2000}. The following {\sc Python} packages were utilized: {\sc NumPy} \citep{numpy}, {\sc SciPy} \citep{scipy}, {\sc Matplotlib} \citep{matplotlib}, {\sc yt} \citep{ytproject}, {\sc ChiantiPy} \citep{Dere2013}. 

\section*{Data Availability}
The data underlying this article will be shared on reasonable
request to the corresponding author. The code used to compute the temperature-dependent cross sections is publicly available at \url{https://github.com/AstroBrandt/XRayCrossSections}.



\bibliographystyle{mnras}
\bibliography{lib} 

\appendix
\section{Cloudy Benchmark Script}\label{sec:cloudApp}
We present the {\sc Cloudy} script which was used for the X-ray benchmarking. The {\sc Cloudy} model consists of a uniform density medium with $n_{\rm H} = 10^3$ solved using the ``sphere'' command. We turn off most induced and grain processes and set the abundances for most metals to zero to better match the methods used in our {\sc Flash} simulations. 
\begin{lstlisting}[float,floatplacement=H, language=bash, caption=Input file for {\sc Cloudy} benchmark]
title XDR source
##radiation sources
CMB
##Source
table SED "plaw.sed"
luminosity 35 range 73.5 to 735 Ryd
radius 16.1938
##Density
hden 3.0
sphere
##Stopping and iterate
stop H2 column density 24
stop temperature linear 3.0
iterate to convergence
##Misc
abundances ISM
no grain qheat
no grain x-ray treatment
no induced processes
no radiation pressure
no scattering opacity
no grain molecules
no line transfer
element carbon abundance -3.853872
element helium abundance -1
element oxygen abundance -3.494850
element silicon abundance -7
element nitrogen off
element sulphur off
element neon off
element aluminium off
element phosphor off
element chlorine off
element argon off
element calcium off
element chromium off
element nickel off
element lithium off
element beryllium off
element fluorine off
element potassium off
element scandium off
element titanium off
element vanadium off
element manganese off
element cobalt off
element copper off
element zinc off
cosmic ray rate -16.523
##output
set save luminosity old
save overview last  "xdr.ovr"
save molecules last "xdr.mol"
save abundances last "xdr.abund"
save continuum last "xdr.cont"
save dr last "xdr.dr"
save PDR last "xdr.pdr"
save grain temperature last "xdr.dtemp"
\end{lstlisting}

\bsp	
\label{lastpage}
\end{document}